\documentclass[preprint,floatfix,superscriptaddress]{revtex4-1}

\usepackage[linktocpage,bookmarksopen,bookmarksnumbered]{hyperref}
\usepackage{graphicx} 
\usepackage{gensymb}
\usepackage{amsmath,bm,amssymb}
\usepackage{amsfonts}
\usepackage[usenames]{color}

\begin{document}


\title{Nonlocal Coulomb interaction and spin-freezing crossover as a route to valence-skipping charge order}

 \author{Siheon Ryee}
 \affiliation{Department of Physics, KAIST, Daejeon 34141, Republic of Korea}
 \affiliation{Condensed Matter Physics and Materials Science Department, Brookhaven National Laboratory, Upton, NY 11973, USA}
 
 \author{Patrick S\'emon}
 \affiliation{Condensed Matter Physics and Materials Science Department, Brookhaven National Laboratory, Upton, NY 11973, USA}
 
 \author{Myung Joon Han}
 \email{mj.han@kaist.ac.kr}
 \affiliation{Department of Physics, KAIST, Daejeon 34141, Republic of Korea}
 
 \author{Sangkook Choi}
 \email{sachoi@bnl.gov}
 \affiliation{Condensed Matter Physics and Materials Science Department, Brookhaven National Laboratory, Upton, NY 11973, USA}
\date{\today}

\begin{abstract}
Multiorbital systems away from global half-filling host intriguing physical properties promoted by Hund's coupling. Despite increasing awareness of this regime dubbed Hund's metal, effect of nonlocal interaction is still elusive.
Here we study a three-orbital model with $1/3$ filling (two electrons per site) including the intersite Coulomb interaction ($V$). Using the $GW$ plus extended dynamical mean-field theory, the valence-skipping charge order transition is shown to be driven by $V$. Most interestingly, the instability to this transition is significantly enhanced in the spin-freezing crossover regime, thereby lowering the critical $V$ to the formation of charge order. This behavior is found to be closely related to the population profile of the atomic multiplet states in the spin-freezing regime. In this regime, maximum spin states are dominant in each total charge subspace with substantial amount of one- and three-electron occupations, which leads to almost equal population of one- and the maximum spin three-electron state.
Our finding unveils another feature of the Hund's metal, and has potential implications for the broad range of multiorbital systems as well as the recently discovered charge order in iron-pnictides.

\end{abstract}
 
\maketitle

\section*{Introduction}
Classifying a number of phases and understanding their relevance to different energy scales has been a central theme of condensed matter physics. In multiorbital systems away from global half-filling, Hund's coupling was shown to promote a bad metallic behavior while simultaneously pushing away the Mott insulating region \cite{Janus,Medici_2011,Georges}. The term Hund's metal \cite{Haule,Yin1} was coined to classify the regime in which the ``Hundness'' not the ``Mottness'' plays a leading role in determining physical properties \cite{Fanfarillo,Stadler2}. The Hund's metal hosts rich phenomena such as finite temperature spin-freezing crossover \cite{Werner1,Nomura,Stadler2}, spin-orbital separation \cite{SOS1,SOS2,SOS3,SOS4,Stadler2,Deng}, anomalous transport behavior \cite{Werner1,Haule,Georges}, increased electronic compressibility \cite{Medici_2017,Arribi}, and the orbital-differentiation \cite{Medici_2011,Bascones,Lanata,Medici_2014,Kostin}. It has been believed to be one of the central doctrines to understand the intriguing physics of (mainly, but not limited to) iron-based superconductors \cite{Haule,Hansmann,Yin1,Yin2,Georges,Bascones,Lanata,nematicity,Medici_2017,Arribi} and ruthenates \cite{Werner1,Mravlje,Georges,Hoshino,SRO}. 

In addition to the above mentioned direct manifestations of Hund's metal regime, its connection and proximity to the symmetry-broken charge-disproportionated phases has recently been highlighted \cite{Isidori,Strand}. Those which are called Hund's insulator \cite{Isidori} and valence-skipping phase \cite{Varma,Harrison,Strand} --- a phase with two different valences while skipping the intermediate one between the two --- are prominent examples. One possible route to the valence-skipping is the negative effective Coulomb repulsion, $U_{\mathrm{eff}} < 0$ \cite{Anderson,Katayama,Strand}.  
Interestingly, a purely intra-atomic origin, namely the anisotropic orbital-multipole scattering, was suggested to be the key ingredient for such valence-skipping phenomena \cite{Strand}.  Furthermore, this phase has potential implications for the electron pairing mechanisms of unconventional superconductivity \cite{Micnas,Strand}.  

The valence-skipping compounds are prevalent in Nature most evidently in the form of charge order (CO) \cite{Strand}. The CO transition has actively been studied in the single-orbital extended Hubbard model presumably in close connection with the superconductivity of cuprates \cite{Keimer}. Notably, as in the case of cuprates, recent experiments reported the CO in the vicinity of the superconducting phase of $A$Fe$_2$As$_2$ ($A=\text{Rb}$, $\text{K}$, $\text{Cs}$), archetypal materials of Hund's metal \cite{Civardi,Wang,Moroni}. Moreover, relevance of charge fluctuations or CO to the superconductivity of iron-pnictides was reported \cite{Zhou}. Thus, it is tempting to presume that the CO is a common ``neighbor'' of unconventional high-temperature superconductivity. On the other hand, one can also envisage the more complexity of the multiorbital CO transition due to the additional energy scales such as Hund's coupling absent in single-orbital models. 

In this work, by employing the state-of-the-art $GW$ plus extended dynamical mean-field theory ($GW+$EDMFT) adapted to multiorbital models, we demonstrate that the valence-skipping CO is driven by intersite nonlocal Coulomb repulsion $V$, and the instability to this phase is significantly enhanced in the spin-freezing crossover regime. This enhancement is shown to be related to the local multiplet population profile. This route to the valence-skipping is distinctive from the anisotropic orbital-multipole scattering mechanism \cite{Strand}.

\section*{Results and Discussion}
We first construct a following model for the two-dimensional square lattice including both local and nonlocal interaction terms:
\begin{align}
\mathcal{H} =& -t\sum_{\langle ij \rangle,\gamma,\sigma}{\big(c^{\dagger}_{i\gamma \sigma}c_{j\gamma \sigma} + \mathrm{H.c.}\big)} - \mu\sum_{i,\gamma,\sigma}{n_{i \gamma \sigma}} \nonumber \\ 
&+ H_{\mathrm{loc}} + H_{\mathrm{nonloc}},
\label{eq1}
\end{align}
where $c^{\dagger}_{i \gamma \sigma}$ ($c_{i \gamma \sigma}$) is the electron creation (annihilation) operator acting on site $i$ with orbital index $\gamma=1,2,3$ and spin index $\sigma=\uparrow, \downarrow$. $t$ ($t>0$) is the hopping amplitude between two nearest-neighbor (NN) sites denoted by $\langle ij \rangle$. We use half-bandwidth $D=4t$ as the unit of energies. $n_{i \gamma \sigma}=c^{\dagger}_{i \gamma \sigma}c_{i \gamma \sigma}$ is the electron number operator. $\mu$ is the chemical potential to be adjusted to obey $1/3$ filling per site; $\sum_{\gamma,\sigma}{\langle n_{i \gamma \sigma} \rangle}=2$. $H_\mathrm{loc}$ is of the Kanamori form containing the onsite Coulomb repulsion $U$, and the Hund's coupling $J$, which reads
\begin{align} 
H_\mathrm{loc} &= U\sum_{i,\gamma,\sigma}{n_{i \gamma \uparrow} n_{i \gamma \downarrow}}
+ (U-2J)\sum_{i,\gamma,\gamma'}^{\gamma \neq \gamma'}{ n_{i \gamma \uparrow} n_{i \gamma' \downarrow}} \nonumber \\
&+ (U-3J)\sum_{i,\gamma,\gamma',\sigma}^{\gamma < \gamma'}{ n_{i \gamma \sigma} n_{i \gamma' \sigma}}  \nonumber \\
&- J\sum_{i,\gamma,\gamma'}^{\gamma \neq \gamma'}{(c^{\dagger}_{i \gamma \uparrow} c_{i \gamma \downarrow} c^{\dagger}_{i \gamma' \downarrow} c_{i \gamma' \uparrow} + c^{\dagger}_{i \gamma \uparrow} c^{\dagger}_{i \gamma \downarrow} c_{i \gamma' \uparrow} c_{i \gamma' \downarrow})}. 
\label{eq2}
\end{align}
$H_\mathrm{nonloc}$ is the interaction term between two NN sites coupled via nonlocal Coulomb repulsion $V$, 
\begin{align}
H_\mathrm{nonloc} &= \sum_{\substack{\langle ij \rangle \\ \gamma,\gamma',\sigma,\sigma'}}{Vn_{i \gamma \sigma}n_{j \gamma' \sigma'}}.
\end{align}

To gain a useful insight for CO transition of the model constructed in Eq.~(\ref{eq1}), we first investigate a simple case of vanishing $t$ and temperature. This simple atomic limit --- a limit where the lattice consists of atoms with zero $t$ among them --- enables us to get analytical solutions, which is found to be a good estimate even under nonzero $t$ and temperature \cite{Strand,Yan,Pawlowski,Kapcia1,Kapcia2}. 
In Fig.~\ref{fig1} we plot the obtained phase diagram (see Supplemental Material A for more details). 
Three different phases are classified according to their valence. We used notation $d^N$ to denote the $N$-electron occupation of a site in the primitive cell. 
Note that the triple point emerges at $V/U=0$ and $J/U=1/3$, which corresponds to the parameter region where the metal resilient to Mott- and Hund's-insulator transition emerges \cite{Isidori}, as well as the valence-skipping phases cease to exist \cite{Strand}. The possible existence of $d^3+d^1$ phase was previously noticed from the slave-boson mean-field by solving the Kanamori Hamiltonian \cite{Isidori}. This state, however, is degenerate at $J/U=1/3$ with $d^2$ and $2d^3+d^0$ phases, and never the ground state unless $V/U>0$. The $2d^3+d^0$ phase is equivalent to the charge-ordered Hund's insulator \cite{Isidori}. 
Note also that other charge orders such as $d^4+d^0$ and $2d^0+d^6$ can be stabilized above the dashed lines depicted in Fig.~\ref{fig1}, which are quite irrelevant for the present study.

At $0 < J/U < 1/3$, we can observe a transition from the isotropic $d^2$ to $d^3+d^1$ valence-skipping CO with ordering wave-vector $(\pi,\pi)$ at the critical $V$ ($V_c$), $V_c = \frac{U}{4}(1-3J/U)$.  
It should be noted that this phase is driven by $V$, not by the anisotropic orbital-multipole scattering since the Kanamori form is free from it by construction \cite{Strand}. 
At $J/U=0$, $V_c$ follows the half-filled single-orbital result of $V_c=U/4$ \cite{Yan,Pawlowski}.

With insight obtained above, we now turn to our $GW+$EDMFT results. The corresponding phase diagram obtained from $GW+$EDMFT is shown in Fig.~\ref{fig2}(a). We identified the CO transition by monitoring the divergence of the static charge susceptibility $\chi(\mathbf{k},i\nu_0)$ ($\nu_0$ is the lowest bosonic matsubara frequency, $\nu_0=0$). The divergence actually occurs at the wave-vector $\mathbf{k}=(\pi,\pi)$ indicating the formation of $d^3+d^1$ order (see Fig.~\ref{fig1} and Fig.~\ref{fig3}(b)). One can also confirm that this CO transition is driven by $V$ (compare Fig.~\ref{fig3}(b) with Fig.~\ref{fig3}(a); see also Supplemental Material C for $\chi(\mathbf{k},i\nu_0)$ at $V=0$).

The actual $GW+$EDMFT results roughly follow the atomic limit estimate at $J/U \leq 0.15$, and are in fair agreement at large $U$ $(U=3,4)$ and $J/U=0.15$.  Even in smaller $U$ region (Fermi-liquid; see Fig.~\ref{fig4}(a)), $GW+$EDMFT results qualitatively follow the atomic limit estimates. This seemingly unusual behavior is attributed to the leading contribution of interaction energy compared to the kinetic energy in determining the CO transition boundary \cite{Terletska}. Note that the Mott phase emerges at $V=0$ for $U=4$ (when $J/U=0.05$) and $U=5$ (when $J/U \leq 0.15$).  
In the current study, we restrict our discussion to the $U$ and $J/U$ region in which the metallic phase is obtained when $V=0$.

At $J/U=0.2$, $GW+$EDMFT results exhibit unprecedented behavior at large $U$ ($U\geq3$): CO instability is significantly enhanced, thereby pushing $V_c$ further below the atomic limit estimates. Notably, the downturn of $V_c$ is most pronounced at $U=4$ followed by a rapid upturn of the phase boundary at $U=5$.
This behavior is not captured either by EDMFT or $GW$ approximations (see Fig.~\ref{fig2}(b) and (c)). On the other hand, at smaller $U$ ($U\leq2$), $V_c$ values obtained from $GW+$EDMFT are almost identical to those of EDMFT, and larger than the atomic limit estimates. The discrepancy between $GW$ results (Fig.~\ref{fig2}(c)) and the others is reasonable since this method cannot properly treat the local physics.
We briefly remark that at further larger $J/U$, especially near $J/U=1/3$ and $V=0$ in which the triple point emerges in Fig.~\ref{fig1}, a signature of the $2d^3+d^0$ phase is also expected. Notably, the presence of this degeneracy point is claimed to play an important role in stabilizing the metallic phase \cite{Isidori}. The triple degeneracy, however, should be lifted by nonzero $V$. We expect that an intriguing physics can happen due to this broken degeneracy, which we leave for future study.

To further illustrate the above intriguing result from $GW+$EDMFT at large $U$ and $J/U$ regime, we investigate the site-resolved charge susceptibility $\chi(\mathbf{R}_i,i\nu_0)=\int{d\mathbf{k}e^{i\mathbf{k}\cdot\mathbf{R}_i}\chi(\mathbf{k},i\nu_0)}$ ($\mathbf{R}_i$ is the position vector of the $i$-th NN). The magnitude of this quantity is enhanced as $V$ increases as shown in Fig.~\ref{fig3}(c) -- (f). Near the CO boundary ($V\simeq 0.9V_c$), the sign of $\chi(\mathbf{R}_i,i\nu_0)$ clearly indicates the CO instability at $\mathbf{k}=(\pi,\pi)$, which has to be plus (minus) for onsite, second, and third (first and fourth) NNs. 

Most interestingly, the large $U$ results exhibit the rapid growth of $\chi(\mathbf{R}_i,i\nu_0)$ as a function of $J/U$ at a finite $V$ (see Fig.~\ref{fig3}(f)). This behavior is in contrast to the smaller $U$ results in which the increase of  $\chi(\mathbf{R}_i,i\nu_0)$ is much more gradual (see Fig.~\ref{fig3}(d)). 
This enhancement of $\chi(\mathbf{R}_i,i\nu_0)$ at $J/U=0.2$ is further manifested by the static effective local interaction, $\mathcal{U}(i\nu_0)$. The intraorbital elements $\mathcal{U}(i\nu_0)_{\gamma\gamma} \equiv \mathcal{U}(i\nu_0)_{\gamma\gamma\gamma\gamma}$ at $U=4$ and $V \simeq 0.9V_c$ shows the large screening effect at $J/U=0.2$; from $\mathcal{U}(i\nu_0)_{\gamma\gamma} = 3.31$ (3.39) at $J/U=0.1$ (0.15) to $\mathcal{U}(i\nu_0)_{\gamma\gamma} = 2.88$ at $J/U=0.2$.
Note also that the substantial amount of nonlocal $\chi(\mathbf{R}_i,i\nu_0)$ exists even at $V=0$ in the larger $U$ and $J/U=0.2$ regime (compare Fig.~\ref{fig3}(c) and (e) and their insets).

Key information for understanding the large enhancement of CO instability is provided by investigating the local self-energy $\Sigma_\mathrm{loc}(i\omega_n)$ ($\omega_n$: fermionic Matsubara frequency). $\Sigma_\mathrm{loc}(i\omega_n)$ shows an interesting behavior near spin-freezing crossover regime \cite{Werner1,Georges}  which is a metal with emerging local moment: large spin susceptibility $\chi_s = \int_{0}^{\beta}{d\tau \langle S_i(\tau)S_i(0) \rangle}$ with substantial dynamical contribution of $\Delta\chi_s = \int_{0}^{\beta}{d\tau \big( \langle S_i(\tau)S_i(0) \rangle - \langle S_i(\beta/2)S_i(0) \rangle \big)}$ \cite{Hoshino}. $S_i=(1/2)\sum_{\gamma}(n_{i\gamma \uparrow}-n_{i\gamma \downarrow})$ is the local spin operator. In this regime, $\mathrm{Im}\Sigma_\mathrm{loc}(i\omega_n)$ is claimed to follow the power-law behavior at low frequency: $\mathrm{Im}\Sigma_\mathrm{loc}(i\omega_n) \simeq -\Gamma + A(\omega_n)^\alpha$ with $\alpha \simeq 0.5$ and $\Gamma \simeq 0$ \cite{Werner1}. Deep inside this crossover where non-FL behavior appears ($\Gamma > 0$ and $\alpha > 0.5$) is called the frozen moment regime \cite{Werner1,Georges,Hoshino}. In Fig.~\ref{fig4}, we summarize our analysis of $\mathrm{Im}\Sigma_\mathrm{loc}(i\omega_n)$.

Fig.~\ref{fig4}(a) shows the correlation between $\alpha$ and $\Delta \chi_s / \chi_s$. By construction, $\Delta \chi_s / \chi_s$ lies in between 0 and 1. The limiting value of $\Delta \chi_s / \chi_s$ indicates either the FL limit when $\Delta \chi_s / \chi_s \to 1$, or the frozen-moment regime when $\Delta \chi_s / \chi_s \to 0$. Thus, we can naturally expect that the spin-freezing regime should lie somewhere in between these two limits. We identify the region of $0.4 \lesssim \alpha \lesssim 0.5$ and $\Gamma \simeq 0$ with the spin-freezing crossover regime. In our parameter range, spin-freezing regime appears for $0.25 < \Delta \chi_s / \chi_s < 0.4$ (see also Supplemental Material D for the correlation of $\alpha$ with $\chi_s^{-1}$ and $\Delta\chi_s$).  
In FL regime, increasing $V$ drives the system to be less correlated. Interestingly at $U=4$ and $J/U=0.1$, $V$ drives the system from the (proximity of) frozen moment to FL and eventually to CO. This behavior can be confirmed by vanishing $\Gamma$ and $\alpha > 0.5$ near $V_c$ (see Fig.~\ref{fig4}(b)).
We also note that EDMFT yields qualitatively similar results (not shown) except that in $U=4$ and $J/U=0.1$, increasing $V$ do not show any signal of transition to the FL.

Most notably, the parameter region showing the unusual downturn of $V_c$ ($U=3,4$ and $J/U=0.2$) corresponds to the spin-freezing crossover regime. As $U$ increases further at $J/U=0.2$, an upturn of the phase boundary appears (see Fig.~\ref{fig2}(a)) as entering deeper into the frozen moment regime. In this range of $U$ and $J/U$, the increasing $V$ tends to reduce $\alpha$ while maintaining $\Gamma=0$ (see Fig.~\ref{fig4}(a) and (c)).
To further clarify the relation between the enhanced CO instability and the spin-freezing crossover, we investigate the local populations (or probabilities) of atomic multiplet states. The  U(1)$_\mathrm{charge}$~$\times$~SU(2)$_\mathrm{spin}$~$\times$~SO(3)$_\mathrm{orbital}$ symmetry of Eq.~(\ref{eq2}) allows us to have the simultaneous eigenstates of charge $N$, orbital $L$, and spin $S$ as $|N,L,S\rangle$ \cite{Janus,Georges,Isidori}. The local population profiles of these eigenstates are plotted in Fig.~\ref{fig5}(a) and (b) as approaching the CO boundary.

One can notice that in spin-freezing crossover regime, maximum $S$ states are dominant in each total charge subspace with substantial amount of $N=1$ and $N=3$ populations (contribution of states other than $N=1,2,3$ subspaces are negligible) (Fig.~\ref{fig5}(c)). It is the effect of $J$ favoring the maximum $S$. Importantly, these $N=1$ and $N=3$ charges are directly related to the $d^3+d^1$ CO phase, which implies the enhanced CO instability in this regime. 
On the other hand in the frozen moment regime, $N=2$ population is more dominant with reduced $N=1$ and $N=3$ portions than the spin-freezing case; compare Fig.~\ref{fig5}(a) and (b) with Fig.~\ref{fig5}(c). 
FL regime exhibits non-negligible excursions to every other $| N,L,S \rangle$ as expected. 
We hereafter denote the population of $|1,1,1/2\rangle$ and $|3,0,3/2\rangle$ by $p_1$ and $p_3$ .

At $U=4$ and $J/U=0.2$ only the maximum $S$ is selected in the $N=3$ subspace, leading to $p_3 \simeq p_1$ (see Fig.~\ref{fig5}(c)). At $U=5$ and $J/U=0.2$ (frozen moment), $p_3 \simeq p_1$ is also found. This case, however, shows more dominant $N=2$ population ($\sim 0.79$ at $V=0$) with reduced $N=1$ and $N=3$ contributions compared to the $U=4$ and $J/U=0.2$ case. In light of this observation, we construct a phenomenological local wave-function $\psi$ consisting of maximum $S$ states, namely $\psi = \sqrt{p_1}|1,1,1/2\rangle + \sqrt{1-2p_1}|2,1,1\rangle + \sqrt{p_1}|3,0,3/2\rangle$ apart from the phase factor which is irrelevant for the evaluation of energy.
The re-calculated $V_c$ estimate (as is done for Fig.~\ref{fig1}) by means of $\psi$ is shown in Fig.~\ref{fig5}(d). We can observe the qualitative agreement with the actual behavior obtained from $GW+$EDMFT at $J/U=0.2$ (see stars in Fig.~\ref{fig5}(d)). This result confirms the role of maximum $S$ states in $N=3$ subspace in enhancing the CO instability. 
This type of interpretation should be valid in large $U$ and $J/U$ limit.
Fig.~\ref{fig5}(d) shows, however, deviations of actual $GW+$EDMFT results at $J/U=0.1$ and $J/U=0.15$. This can be attributed to the non-negligible amount smaller $S$ states in $N=3$ subspace, and the fundamental inadequacy of this kind of approach for FL regime.

In conclusion, we have shown by employing $GW+$EDMFT that in spin-freezing regime, significant enhancement of $d^3+d^1$ CO instability appears. This enhancement is found to be closely related to the local multiplet population profile: maximum spin states are dominant in each total charge subspace with substantial amount of $N=1$ and $N=3$ occupations. 
The observed $d^3+d^1$ CO transition is driven by $V$, and is also a distinctive route from the anisotropic orbital-multipole scattering mechanism to the valence-skipping phenomena \cite{Strand}. Our study
unveils another feature of the Hund's metal, and has potential implications for other multiorbital systems and observed CO in Hund's metal $A$Fe$_2$As$_2$ ($A=\text{Rb}$, $\text{K}$, $\text{Cs}$) \cite{Civardi,Wang,Moroni}.

\section*{Methods}
$GW+$EDMFT is derivable from the $\mathrm{\Psi}[G,W]$ functional ($G$: Green's function, $W$: fully screened Coulomb interaction) \cite{Almbladh} as $\mathrm{\Psi}^{GW+\mathrm{EDMFT}}[G,W]=\mathrm{\Psi}^{\mathrm{EDMFT}}[G_\mathrm{loc},W_\mathrm{loc}] + \mathrm{\Psi}^{GW}_\mathrm{nonloc}[G,W]$, where EDMFT is supplemented with nonlocal $GW$ functional \cite{Sun,Chitra,Biermann,Tomczak,Ayral}. This approach allows a nonperturbative solution of the auxiliary impurity model with self-consistently determined local fermionic and bosonic Weiss fields. The bosonic Weiss field $\mathcal{U}(i\nu_n)$ ($\nu_n$: bosonic Matsubara frequency) is the effective impurity interaction whose value is renormalized by dynamical screening effect. The importance of this effect has recently been highlighted \cite{Ayral_prl,Hansmann2,Ayral,Huang,Ayral2,Boehnke,multitier,Golez}.
We performed calculations within the paramagnetic isotropic phase and inverse temperature of $\beta D=100$. An impurity model was solved using the \textsc{ComCTQMC} implementation \cite{Choi} of the hybridization-expansion CTQMC algorithm \cite{CTQMC,CTQMC2}. 
Both local and nonlocal interaction terms were decoupled via Hubbard-Stratonovich transformation to treat them on an equal footing \cite{Ayral}. In our current implementation, due to the computational complexity we measured only the density-density type of two-particle correlation functions from the impurity; $\chi_\mathrm{imp}(\tau)=\langle \mathcal{T}_{\tau} n_{\gamma \sigma}(\tau)n_{\gamma' \sigma'}(0) \rangle$ ($\tau$: imaginary time).  
The non-density-density type functions are responsible for the screening of non-monopole terms of charge distribution, making our approximation physically reasonable since these terms are ill-screened \cite{multitier}. All three methods ($GW+$EDMFT, EDMFT, and $GW$) were performed self-consistently. See Supplemental Material B for further details of our $GW+$EDMFT calculations.

\section*{Data Availability}
The data that support the findings of this study are available from corresponding authors upon reasonable request.    

\section*{Code Availability}
The computer code used for this study is available upon reasonable request.  
 
\section*{Acknowledgements}
S.R. thanks T. Ayral and J.-H. Sim for useful conversation at the early stage of this work. 
S.R. and M.J.H. were supported by BK21plus program, Basic Science Research Program (2018R1A2B2005204) and Creative Materials Discovery Program through NRF (2018M3D1A1058754).
P.S. and S.C. were supported by the U.S. Department of Energy, Office of Science, Basic Energy Sciences as a part of the Computational Materials Science Program. 
This research used resources of the National Energy Research Scientific Computing Center (NERSC), a U.S. Department of Energy Office of Science User Facility operated under Contract No. DE-AC02-05CH11231.

\section*{Competing Interests}
The authors declare no competing financial or non-financial interests. 


\section*{Author Contributions}
S.C. conceived the project. S.R. developed the code on top of quantum impurity solver built by P.S. S.R. performed all calculations. S.C., S.R., and M.J.H. discussed the data and wrote the manuscript. All authors commented on the document.



\bibliography{ref}

\begin{thebibliography}{10}
\expandafter\ifx\csname url\endcsname\relax
  \def\url#1{\texttt{#1}}\fi
\expandafter\ifx\csname urlprefix\endcsname\relax\def\urlprefix{URL }\fi
\providecommand{\bibinfo}[2]{#2}
\providecommand{\eprint}[2][]{\url{#2}}

\bibitem{Janus}
\bibinfo{author}{de' Medici, L.}, \bibinfo{author}{Mravlje, J.} \&
  \bibinfo{author}{Georges, A.}
\newblock \bibinfo{title}{Janus-faced influence of $\mathrm{H}$und's rule
  coupling in strongly correlated materials}.
\newblock \emph{\bibinfo{journal}{Phys. Rev. Lett.}}
  \textbf{\bibinfo{volume}{107}}, \bibinfo{pages}{256401 -- 256404}
  (\bibinfo{year}{2011}).
\newblock
  \urlprefix\url{https://link.aps.org/doi/10.1103/PhysRevLett.107.256401}.

\bibitem{Medici_2011}
\bibinfo{author}{de' Medici, L.}
\newblock \bibinfo{title}{Hund's coupling and its key role in tuning
  multiorbital correlations}.
\newblock \emph{\bibinfo{journal}{Phys. Rev. B}} \textbf{\bibinfo{volume}{83}},
  \bibinfo{pages}{205112 -- 205122} (\bibinfo{year}{2011}).
\newblock \urlprefix\url{https://link.aps.org/doi/10.1103/PhysRevB.83.205112}.

\bibitem{Georges}
\bibinfo{author}{Georges, A.}, \bibinfo{author}{de~Medici, L.} \&
  \bibinfo{author}{Mravlje, J.}
\newblock \bibinfo{title}{Strong correlations from $\mathrm{Hund}$'s coupling}.
\newblock \emph{\bibinfo{journal}{Annu. Rev. Condens. Matter Phys.}}
  \textbf{\bibinfo{volume}{4}}, \bibinfo{pages}{137 -- 178}
  (\bibinfo{year}{2013}).

\bibitem{Haule}
\bibinfo{author}{Haule, K.} \& \bibinfo{author}{Kotliar, G.}
\newblock \bibinfo{title}{Coherence--incoherence crossover in the normal state
  of iron oxypnictides and importance of $\mathrm{Hund}$'s rule coupling}.
\newblock \emph{\bibinfo{journal}{New Journal of Physics}}
  \textbf{\bibinfo{volume}{11}}, \bibinfo{pages}{025021 -- 025033}
  (\bibinfo{year}{2009}).
\newblock
  \urlprefix\url{https://doi.org/10.1088%2F1367-2630%2F11%2F2%2F025021}.

\bibitem{Yin1}
\bibinfo{author}{Yin, Z.}, \bibinfo{author}{Haule, K.} \&
  \bibinfo{author}{Kotliar, G.}
\newblock \bibinfo{title}{Kinetic frustration and the nature of the magnetic
  and paramagnetic states in iron pnictides and iron chalcogenides}.
\newblock \emph{\bibinfo{journal}{Nature Materials}}
  \textbf{\bibinfo{volume}{10}}, \bibinfo{pages}{932 -- 935}
  (\bibinfo{year}{2011}).

\bibitem{Fanfarillo}
\bibinfo{author}{Fanfarillo, L.} \& \bibinfo{author}{Bascones, E.}
\newblock \bibinfo{title}{Electronic correlations in $\mathrm{Hund}$ metals}.
\newblock \emph{\bibinfo{journal}{Phys. Rev. B}} \textbf{\bibinfo{volume}{92}},
  \bibinfo{pages}{075136 -- 075142} (\bibinfo{year}{2015}).
\newblock \urlprefix\url{https://link.aps.org/doi/10.1103/PhysRevB.92.075136}.

\bibitem{Stadler2}
\bibinfo{author}{Stadler, K.}, \bibinfo{author}{Kotliar, G.},
  \bibinfo{author}{Weichselbaum, A.} \& \bibinfo{author}{von Delft, J.}
\newblock \bibinfo{title}{$\mathrm{Hundness}$ versus $\mathrm{Mottness}$ in a
  three-band $\mathrm{Hubbard}$–$\mathrm{Hund}$ model: on the origin of
  strong correlations in $\mathrm{Hund}$ metals}.
\newblock \emph{\bibinfo{journal}{Annals of Physics}}
  \textbf{\bibinfo{volume}{405}}, \bibinfo{pages}{365 -- 409}
  (\bibinfo{year}{2019}).
\newblock
  \urlprefix\url{http://www.sciencedirect.com/science/article/pii/S0003491618302793}.

\bibitem{Werner1}
\bibinfo{author}{Werner, P.}, \bibinfo{author}{Gull, E.},
  \bibinfo{author}{Troyer, M.} \& \bibinfo{author}{Millis, A.~J.}
\newblock \bibinfo{title}{Spin freezing transition and
  non-$\mathrm{Fermi}$-liquid self-energy in a three-orbital model}.
\newblock \emph{\bibinfo{journal}{Phys. Rev. Lett.}}
  \textbf{\bibinfo{volume}{101}}, \bibinfo{pages}{166405 -- 166408}
  (\bibinfo{year}{2008}).
\newblock
  \urlprefix\url{https://link.aps.org/doi/10.1103/PhysRevLett.101.166405}.

\bibitem{Nomura}
\bibinfo{author}{Nomura, Y.}, \bibinfo{author}{Sakai, S.} \&
  \bibinfo{author}{Arita, R.}
\newblock \bibinfo{title}{Nonlocal correlations induced by hund's coupling: a
  cluster $\mathrm{DMFT}$ study}.
\newblock \emph{\bibinfo{journal}{Phys. Rev. B}} \textbf{\bibinfo{volume}{91}},
  \bibinfo{pages}{235107 -- 235111} (\bibinfo{year}{2015}).
\newblock \urlprefix\url{https://link.aps.org/doi/10.1103/PhysRevB.91.235107}.

\bibitem{SOS1}
\bibinfo{author}{Yin, Z.~P.}, \bibinfo{author}{Haule, K.} \&
  \bibinfo{author}{Kotliar, G.}
\newblock \bibinfo{title}{Fractional power-law behavior and its origin in
  iron-chalcogenide and ruthenate superconductors: Insights from
  first-principles calculations}.
\newblock \emph{\bibinfo{journal}{Phys. Rev. B}} \textbf{\bibinfo{volume}{86}},
  \bibinfo{pages}{195141 -- 195149} (\bibinfo{year}{2012}).
\newblock \urlprefix\url{https://link.aps.org/doi/10.1103/PhysRevB.86.195141}.

\bibitem{SOS2}
\bibinfo{author}{Horvat, A.}, \bibinfo{author}{\ifmmode~\check{Z}\else
  \v{Z}\fi{}itko, R.} \& \bibinfo{author}{Mravlje, J.}
\newblock \bibinfo{title}{Low-energy physics of three-orbital impurity model
  with $\mathrm{Kanamori}$ interaction}.
\newblock \emph{\bibinfo{journal}{Phys. Rev. B}} \textbf{\bibinfo{volume}{94}},
  \bibinfo{pages}{165140 -- 165150} (\bibinfo{year}{2016}).
\newblock \urlprefix\url{https://link.aps.org/doi/10.1103/PhysRevB.94.165140}.

\bibitem{SOS3}
\bibinfo{author}{Aron, C.} \& \bibinfo{author}{Kotliar, G.}
\newblock \bibinfo{title}{Analytic theory of $\mathrm{Hund}$'s metals: a
  renormalization group perspective}.
\newblock \emph{\bibinfo{journal}{Phys. Rev. B}} \textbf{\bibinfo{volume}{91}},
  \bibinfo{pages}{041110 -- 041114} (\bibinfo{year}{2015}).
\newblock \urlprefix\url{https://link.aps.org/doi/10.1103/PhysRevB.91.041110}.

\bibitem{SOS4}
\bibinfo{author}{Horvat, A.}, \bibinfo{author}{Zitko, R.} \&
  \bibinfo{author}{Mravlje, J.}
\newblock \bibinfo{title}{Non-$\mathrm{Fermi}$-liquid fixed point in
  multi-orbital $\mathrm{Kondo}$ impurity model relevant for $\mathrm{Hund}$'s
  metals}.
\newblock \emph{\bibinfo{journal}{arXiv preprint arXiv:1907.07100}}
  (\bibinfo{year}{2019}).

\bibitem{Deng}
\bibinfo{author}{Deng, X.} \emph{et~al.}
\newblock \bibinfo{title}{Signatures of $\mathrm{M}$ottness and
  $\mathrm{H}$undness in archetypal correlated metals}.
\newblock \emph{\bibinfo{journal}{Nature Communications}}
  \textbf{\bibinfo{volume}{10}}, \bibinfo{pages}{2721 -- 2730}
  (\bibinfo{year}{2019}).

\bibitem{Medici_2017}
\bibinfo{author}{de' Medici, L.}
\newblock \bibinfo{title}{Hund's induced $\mathrm{Fermi}$-liquid instabilities
  and enhanced quasiparticle interactions}.
\newblock \emph{\bibinfo{journal}{Phys. Rev. Lett.}}
  \textbf{\bibinfo{volume}{118}}, \bibinfo{pages}{167003 -- 167007}
  (\bibinfo{year}{2017}).
\newblock
  \urlprefix\url{https://link.aps.org/doi/10.1103/PhysRevLett.118.167003}.

\bibitem{Arribi}
\bibinfo{author}{Villar~Arribi, P.} \& \bibinfo{author}{de' Medici, L.}
\newblock \bibinfo{title}{Hund-enhanced electronic compressibility in
  $\mathrm{Fe}\mathrm{Se}$ and its correlation with $\mathit{T}_c$}.
\newblock \emph{\bibinfo{journal}{Phys. Rev. Lett.}}
  \textbf{\bibinfo{volume}{121}}, \bibinfo{pages}{197001 -- 197006}
  (\bibinfo{year}{2018}).
\newblock
  \urlprefix\url{https://link.aps.org/doi/10.1103/PhysRevLett.121.197001}.

\bibitem{Bascones}
\bibinfo{author}{Bascones, E.}, \bibinfo{author}{Valenzuela, B.} \&
  \bibinfo{author}{Calder\'on, M.~J.}
\newblock \bibinfo{title}{Orbital differentiation and the role of orbital
  ordering in the magnetic state of $\mathrm{Fe}$ superconductors}.
\newblock \emph{\bibinfo{journal}{Phys. Rev. B}} \textbf{\bibinfo{volume}{86}},
  \bibinfo{pages}{174508 -- 174513} (\bibinfo{year}{2012}).
\newblock \urlprefix\url{https://link.aps.org/doi/10.1103/PhysRevB.86.174508}.

\bibitem{Lanata}
\bibinfo{author}{Lanat\`a, N.} \emph{et~al.}
\newblock \bibinfo{title}{Orbital selectivity in $\mathrm{Hund}$'s metals: the
  iron chalcogenides}.
\newblock \emph{\bibinfo{journal}{Phys. Rev. B}} \textbf{\bibinfo{volume}{87}},
  \bibinfo{pages}{045122 -- 045126} (\bibinfo{year}{2013}).
\newblock \urlprefix\url{https://link.aps.org/doi/10.1103/PhysRevB.87.045122}.

\bibitem{Medici_2014}
\bibinfo{author}{de' Medici, L.}, \bibinfo{author}{Giovannetti, G.} \&
  \bibinfo{author}{Capone, M.}
\newblock \bibinfo{title}{Selective $\mathrm{Mott}$ physics as a key to iron
  superconductors}.
\newblock \emph{\bibinfo{journal}{Phys. Rev. Lett.}}
  \textbf{\bibinfo{volume}{112}}, \bibinfo{pages}{177001 -- 177005}
  (\bibinfo{year}{2014}).
\newblock
  \urlprefix\url{https://link.aps.org/doi/10.1103/PhysRevLett.112.177001}.

\bibitem{Kostin}
\bibinfo{author}{Kostin, A.} \emph{et~al.}
\newblock \bibinfo{title}{Imaging orbital-selective quasiparticles in the
  $\mathrm{H}$und’s metal state of $\mathrm{Fe}\mathrm{Se}$}.
\newblock \emph{\bibinfo{journal}{Nature Materias}}
  \textbf{\bibinfo{volume}{17}}, \bibinfo{pages}{869 -- 874}
  (\bibinfo{year}{2018}).

\bibitem{Hansmann}
\bibinfo{author}{Hansmann, P.} \emph{et~al.}
\newblock \bibinfo{title}{Dichotomy between large local and small ordered
  magnetic moments in iron-based superconductors}.
\newblock \emph{\bibinfo{journal}{Phys. Rev. Lett.}}
  \textbf{\bibinfo{volume}{104}}, \bibinfo{pages}{197002 -- 197005}
  (\bibinfo{year}{2010}).
\newblock
  \urlprefix\url{https://link.aps.org/doi/10.1103/PhysRevLett.104.197002}.

\bibitem{Yin2}
\bibinfo{author}{Yin, Z.}, \bibinfo{author}{Haule, K.} \&
  \bibinfo{author}{Kotliar, G.}
\newblock \bibinfo{title}{Magnetism and charge dynamics in iron pnictides}.
\newblock \emph{\bibinfo{journal}{Nature Physics}}
  \textbf{\bibinfo{volume}{7}}, \bibinfo{pages}{294 -- 297}
  (\bibinfo{year}{2011}).

\bibitem{nematicity}
\bibinfo{author}{Fanfarillo, L.}, \bibinfo{author}{Giovannetti, G.},
  \bibinfo{author}{Capone, M.} \& \bibinfo{author}{Bascones, E.}
\newblock \bibinfo{title}{Nematicity at the $\mathrm{H}$und's metal crossover
  in iron superconductors}.
\newblock \emph{\bibinfo{journal}{Phys. Rev. B}} \textbf{\bibinfo{volume}{95}},
  \bibinfo{pages}{144511 -- 144517} (\bibinfo{year}{2017}).
\newblock \urlprefix\url{https://link.aps.org/doi/10.1103/PhysRevB.95.144511}.

\bibitem{Mravlje}
\bibinfo{author}{Mravlje, J.} \emph{et~al.}
\newblock \bibinfo{title}{Coherence-incoherence crossover and the
  mass-renormalization puzzles in $\mathrm{Sr}_2\mathrm{RuO}_4$}.
\newblock \emph{\bibinfo{journal}{Phys. Rev. Lett.}}
  \textbf{\bibinfo{volume}{106}}, \bibinfo{pages}{096401 -- 096404}
  (\bibinfo{year}{2011}).
\newblock
  \urlprefix\url{https://link.aps.org/doi/10.1103/PhysRevLett.106.096401}.

\bibitem{Hoshino}
\bibinfo{author}{Hoshino, S.} \& \bibinfo{author}{Werner, P.}
\newblock \bibinfo{title}{Superconductivity from emerging magnetic moments}.
\newblock \emph{\bibinfo{journal}{Phys. Rev. Lett.}}
  \textbf{\bibinfo{volume}{115}}, \bibinfo{pages}{247001 -- 247005}
  (\bibinfo{year}{2015}).
\newblock
  \urlprefix\url{https://link.aps.org/doi/10.1103/PhysRevLett.115.247001}.

\bibitem{SRO}
\bibinfo{author}{Mravlje, J.} \& \bibinfo{author}{Georges, A.}
\newblock \bibinfo{title}{Thermopower and entropy: Lessons from
  $\mathrm{Sr}_2\mathrm{Ru}\mathrm{O}_4$}.
\newblock \emph{\bibinfo{journal}{Phys. Rev. Lett.}}
  \textbf{\bibinfo{volume}{117}}, \bibinfo{pages}{036401 -- 036405}
  (\bibinfo{year}{2016}).
\newblock
  \urlprefix\url{https://link.aps.org/doi/10.1103/PhysRevLett.117.036401}.

\bibitem{Isidori}
\bibinfo{author}{Isidori, A.} \emph{et~al.}
\newblock \bibinfo{title}{Charge disproportionation, mixed valence, and
  $\mathrm{Janus}$ effect in multiorbital systems: A tale of two insulators}.
\newblock \emph{\bibinfo{journal}{Phys. Rev. Lett.}}
  \textbf{\bibinfo{volume}{122}}, \bibinfo{pages}{186401 -- 186406}
  (\bibinfo{year}{2019}).
\newblock
  \urlprefix\url{https://link.aps.org/doi/10.1103/PhysRevLett.122.186401}.

\bibitem{Strand}
\bibinfo{author}{Strand, H. U.~R.}
\newblock \bibinfo{title}{Valence-skipping and negative-$\mathit{U}$ in the
  $\mathit{d}$-band from repulsive local $\mathrm{Coulomb}$ interaction}.
\newblock \emph{\bibinfo{journal}{Phys. Rev. B}} \textbf{\bibinfo{volume}{90}},
  \bibinfo{pages}{155108 -- 155113} (\bibinfo{year}{2014}).
\newblock \urlprefix\url{https://link.aps.org/doi/10.1103/PhysRevB.90.155108}.

\bibitem{Varma}
\bibinfo{author}{Varma, C.~M.}
\newblock \bibinfo{title}{Missing valence states, diamagnetic insulators, and
  superconductors}.
\newblock \emph{\bibinfo{journal}{Phys. Rev. Lett.}}
  \textbf{\bibinfo{volume}{61}}, \bibinfo{pages}{2713 -- 2716}
  (\bibinfo{year}{1988}).
\newblock \urlprefix\url{https://link.aps.org/doi/10.1103/PhysRevLett.61.2713}.

\bibitem{Harrison}
\bibinfo{author}{Harrison, W.~A.}
\newblock \bibinfo{title}{Valence-skipping compounds as positive-$\mathrm{U}$
  electronic systems}.
\newblock \emph{\bibinfo{journal}{Phys. Rev. B}} \textbf{\bibinfo{volume}{74}},
  \bibinfo{pages}{245128 -- 245131} (\bibinfo{year}{2006}).
\newblock \urlprefix\url{https://link.aps.org/doi/10.1103/PhysRevB.74.245128}.

\bibitem{Anderson}
\bibinfo{author}{Anderson, P.~W.}
\newblock \bibinfo{title}{Model for the electronic structure of amorphous
  semiconductors}.
\newblock \emph{\bibinfo{journal}{Phys. Rev. Lett.}}
  \textbf{\bibinfo{volume}{34}}, \bibinfo{pages}{953 -- 955}
  (\bibinfo{year}{1975}).
\newblock \urlprefix\url{https://link.aps.org/doi/10.1103/PhysRevLett.34.953}.

\bibitem{Katayama}
\bibinfo{author}{Katayama-Yoshida, H.} \& \bibinfo{author}{Zunger, A.}
\newblock \bibinfo{title}{Exchange-correlation-induced negative effective
  $\mathit{U}$}.
\newblock \emph{\bibinfo{journal}{Phys. Rev. Lett.}}
  \textbf{\bibinfo{volume}{55}}, \bibinfo{pages}{1618 -- 1621}
  (\bibinfo{year}{1985}).
\newblock \urlprefix\url{https://link.aps.org/doi/10.1103/PhysRevLett.55.1618}.

\bibitem{Micnas}
\bibinfo{author}{Micnas, R.}, \bibinfo{author}{Ranninger, J.} \&
  \bibinfo{author}{Robaszkiewicz, S.}
\newblock \bibinfo{title}{Superconductivity in narrow-band systems with local
  nonretarded attractive interactions}.
\newblock \emph{\bibinfo{journal}{Rev. Mod. Phys.}}
  \textbf{\bibinfo{volume}{62}}, \bibinfo{pages}{113 -- 171}
  (\bibinfo{year}{1990}).
\newblock \urlprefix\url{https://link.aps.org/doi/10.1103/RevModPhys.62.113}.

\bibitem{Keimer}
\bibinfo{author}{Keimer, B.}, \bibinfo{author}{Kivelson, S.~A.},
  \bibinfo{author}{Norman, M.~R.}, \bibinfo{author}{Uchida, S.} \&
  \bibinfo{author}{Zaanen, J.}
\newblock \bibinfo{title}{From quantum matter to high-temperature
  superconductivity in copper oxides}.
\newblock \emph{\bibinfo{journal}{Nature}} \textbf{\bibinfo{volume}{518}},
  \bibinfo{pages}{179 -- 186} (\bibinfo{year}{2015}).

\bibitem{Civardi}
\bibinfo{author}{Civardi, E.}, \bibinfo{author}{Moroni, M.},
  \bibinfo{author}{Babij, M.}, \bibinfo{author}{Bukowski, Z.} \&
  \bibinfo{author}{Carretta, P.}
\newblock \bibinfo{title}{Superconductivity emerging from an electronic phase
  separation in the charge ordered phase of
  $\mathrm{Rb}\mathrm{Fe}_2\mathrm{As}_2$}.
\newblock \emph{\bibinfo{journal}{Phys. Rev. Lett.}}
  \textbf{\bibinfo{volume}{117}}, \bibinfo{pages}{217001 -- 217006}
  (\bibinfo{year}{2016}).
\newblock
  \urlprefix\url{https://link.aps.org/doi/10.1103/PhysRevLett.117.217001}.

\bibitem{Wang}
\bibinfo{author}{Wang, P.~S.} \emph{et~al.}
\newblock \bibinfo{title}{Nearly critical spin and charge fluctuations in
  $\mathrm{KFe}_{2}\mathrm{As}_2$ observed by high-pressure $\mathrm{NMR}$}.
\newblock \emph{\bibinfo{journal}{Phys. Rev. B}} \textbf{\bibinfo{volume}{93}},
  \bibinfo{pages}{085129 -- 085135} (\bibinfo{year}{2016}).
\newblock \urlprefix\url{https://link.aps.org/doi/10.1103/PhysRevB.93.085129}.

\bibitem{Moroni}
\bibinfo{author}{Moroni, M.} \emph{et~al.}
\newblock \bibinfo{title}{Charge and nematic orders in
  $\mathrm{A}\mathrm{Fe}_2\mathrm{As}_2$ $(\mathrm{A}=\mathrm{Rb},\mathrm{Cs})$
  superconductors}.
\newblock \emph{\bibinfo{journal}{Phys. Rev. B}} \textbf{\bibinfo{volume}{99}},
  \bibinfo{pages}{235147 -- 235153} (\bibinfo{year}{2019}).
\newblock \urlprefix\url{https://link.aps.org/doi/10.1103/PhysRevB.99.235147}.

\bibitem{Zhou}
\bibinfo{author}{Zhou, S.}, \bibinfo{author}{Kotliar, G.} \&
  \bibinfo{author}{Wang, Z.}
\newblock \bibinfo{title}{Extended $\mathrm{Hubbard}$ model of
  superconductivity driven by charge fluctuations in iron pnictides}.
\newblock \emph{\bibinfo{journal}{Phys. Rev. B}} \textbf{\bibinfo{volume}{84}},
  \bibinfo{pages}{140505(R) -- 140509} (\bibinfo{year}{2011}).
\newblock \urlprefix\url{https://link.aps.org/doi/10.1103/PhysRevB.84.140505}.

\bibitem{Yan}
\bibinfo{author}{Yan, X.-Z.}
\newblock \bibinfo{title}{Theory of the extended $\mathrm{Hubbard}$ model at
  half filling}.
\newblock \emph{\bibinfo{journal}{Phys. Rev. B}} \textbf{\bibinfo{volume}{48}},
  \bibinfo{pages}{7140 -- 7147} (\bibinfo{year}{1993}).
\newblock \urlprefix\url{https://link.aps.org/doi/10.1103/PhysRevB.48.7140}.

\bibitem{Pawlowski}
\bibinfo{author}{Paw{\l}owski, G.}
\newblock \bibinfo{title}{Charge orderings in the atomic limit of the extended
  $\mathrm{H}$ubbard model}.
\newblock \emph{\bibinfo{journal}{The European Physical Journal B - Condensed
  Matter and Complex Systems}} \textbf{\bibinfo{volume}{53}},
  \bibinfo{pages}{471 -- 479} (\bibinfo{year}{2006}).
\newblock \urlprefix\url{https://doi.org/10.1140/epjb/e2006-00409-1}.

\bibitem{Kapcia1}
\bibinfo{author}{Kapcia, K.~J.}, \bibinfo{author}{Robaszkiewicz, S.},
  \bibinfo{author}{Capone, M.} \& \bibinfo{author}{Amaricci, A.}
\newblock \bibinfo{title}{Doping-driven metal-insulator transitions and charge
  orderings in the extended $\mathrm{Hubbard}$ model}.
\newblock \emph{\bibinfo{journal}{Phys. Rev. B}} \textbf{\bibinfo{volume}{95}},
  \bibinfo{pages}{125112 -- 125122} (\bibinfo{year}{2017}).
\newblock \urlprefix\url{https://link.aps.org/doi/10.1103/PhysRevB.95.125112}.

\bibitem{Kapcia2}
\bibinfo{author}{Kapcia, K.~J.}, \bibinfo{author}{Bara\ifmmode~\acute{n}\else
  \'{n}\fi{}ski, J.} \& \bibinfo{author}{Ptok, A.}
\newblock \bibinfo{title}{Diversity of charge orderings in correlated systems}.
\newblock \emph{\bibinfo{journal}{Phys. Rev. E}} \textbf{\bibinfo{volume}{96}},
  \bibinfo{pages}{042104 -- 042115} (\bibinfo{year}{2017}).
\newblock \urlprefix\url{https://link.aps.org/doi/10.1103/PhysRevE.96.042104}.

\bibitem{Terletska}
\bibinfo{author}{Terletska, H.}, \bibinfo{author}{Chen, T.} \&
  \bibinfo{author}{Gull, E.}
\newblock \bibinfo{title}{Charge ordering and correlation effects in the
  extended $\mathrm{Hubbard}$ model}.
\newblock \emph{\bibinfo{journal}{Phys. Rev. B}} \textbf{\bibinfo{volume}{95}},
  \bibinfo{pages}{115149 -- 115159} (\bibinfo{year}{2017}).
\newblock \urlprefix\url{https://link.aps.org/doi/10.1103/PhysRevB.95.115149}.

\bibitem{Almbladh}
\bibinfo{author}{Almbladh, C.-O.}, \bibinfo{author}{von Barth, U.} \&
  \bibinfo{author}{van Leeuwen, R.}
\newblock \bibinfo{title}{Variational total energies from $\mathit{\Phi}$- and
  $\mathit{\Psi}$- derivable theories}.
\newblock \emph{\bibinfo{journal}{International Journal of Modern Physics B}}
  \textbf{\bibinfo{volume}{13}}, \bibinfo{pages}{535 -- 541}
  (\bibinfo{year}{1999}).
\newblock \urlprefix\url{https://doi.org/10.1142/S0217979299000436}.

\bibitem{Sun}
\bibinfo{author}{Sun, P.} \& \bibinfo{author}{Kotliar, G.}
\newblock \bibinfo{title}{Extended dynamical mean-field theory and
  $\mathrm{GW}$ method}.
\newblock \emph{\bibinfo{journal}{Phys. Rev. B}} \textbf{\bibinfo{volume}{66}},
  \bibinfo{pages}{085120 -- 085139} (\bibinfo{year}{2002}).
\newblock \urlprefix\url{https://link.aps.org/doi/10.1103/PhysRevB.66.085120}.

\bibitem{Chitra}
\bibinfo{author}{Chitra, R.} \& \bibinfo{author}{Kotliar, G.}
\newblock \bibinfo{title}{Effective-action approach to strongly correlated
  fermion systems}.
\newblock \emph{\bibinfo{journal}{Phys. Rev. B}} \textbf{\bibinfo{volume}{63}},
  \bibinfo{pages}{115110 -- 115118} (\bibinfo{year}{2001}).
\newblock \urlprefix\url{https://link.aps.org/doi/10.1103/PhysRevB.63.115110}.

\bibitem{Biermann}
\bibinfo{author}{Biermann, S.}, \bibinfo{author}{Aryasetiawan, F.} \&
  \bibinfo{author}{Georges, A.}
\newblock \bibinfo{title}{First-principles approach to the electronic structure
  of strongly correlated systems: combining the $\mathit{GW}$ approximation and
  dynamical nean-field theory}.
\newblock \emph{\bibinfo{journal}{Phys. Rev. Lett.}}
  \textbf{\bibinfo{volume}{90}}, \bibinfo{pages}{086402 -- 086405}
  (\bibinfo{year}{2003}).
\newblock
  \urlprefix\url{https://link.aps.org/doi/10.1103/PhysRevLett.90.086402}.

\bibitem{Tomczak}
\bibinfo{author}{Tomczak, J.~M.}, \bibinfo{author}{Casula, M.},
  \bibinfo{author}{Miyake, T.} \& \bibinfo{author}{Biermann, S.}
\newblock \bibinfo{title}{Asymmetry in band widening and quasiparticle
  lifetimes in $\mathrm{Sr}\mathrm{V}\mathrm{O}_3$: Competition between
  screened exchange and local correlations from combined $\mathit{GW}$ and
  dynamical mean-field theory $\mathit{GW}$ + $\mathrm{DMFT}$}.
\newblock \emph{\bibinfo{journal}{Phys. Rev. B}} \textbf{\bibinfo{volume}{90}},
  \bibinfo{pages}{165138 -- 165161} (\bibinfo{year}{2014}).
\newblock \urlprefix\url{https://link.aps.org/doi/10.1103/PhysRevB.90.165138}.

\bibitem{Ayral}
\bibinfo{author}{Ayral, T.}, \bibinfo{author}{Biermann, S.} \&
  \bibinfo{author}{Werner, P.}
\newblock \bibinfo{title}{Screening and nonlocal correlations in the extended
  hubbard model from self-consistent combined $\mathit{GW}$ and dynamical mean
  field theory}.
\newblock \emph{\bibinfo{journal}{Phys. Rev. B}} \textbf{\bibinfo{volume}{87}},
  \bibinfo{pages}{125149 -- 125169} (\bibinfo{year}{2013}).
\newblock \urlprefix\url{https://link.aps.org/doi/10.1103/PhysRevB.87.125149}.

\bibitem{Ayral_prl}
\bibinfo{author}{Ayral, T.}, \bibinfo{author}{Werner, P.} \&
  \bibinfo{author}{Biermann, S.}
\newblock \bibinfo{title}{Spectral properties of correlated materials: local
  vertex and nonlocal two-particle correlations from combined $\mathit{GW}$ and
  dynamical nean field theory}.
\newblock \emph{\bibinfo{journal}{Phys. Rev. Lett.}}
  \textbf{\bibinfo{volume}{109}}, \bibinfo{pages}{226401 -- 226405}
  (\bibinfo{year}{2012}).
\newblock
  \urlprefix\url{https://link.aps.org/doi/10.1103/PhysRevLett.109.226401}.

\bibitem{Hansmann2}
\bibinfo{author}{Hansmann, P.}, \bibinfo{author}{Ayral, T.},
  \bibinfo{author}{Vaugier, L.}, \bibinfo{author}{Werner, P.} \&
  \bibinfo{author}{Biermann, S.}
\newblock \bibinfo{title}{Long-range $\mathrm{C}$oulomb interactions in surface
  systems: a first-principles description within self-consistently combined
  $\mathit{GW}$ and dynamical mean-field theory}.
\newblock \emph{\bibinfo{journal}{Phys. Rev. Lett.}}
  \textbf{\bibinfo{volume}{110}}, \bibinfo{pages}{166401 -- 166405}
  (\bibinfo{year}{2013}).
\newblock
  \urlprefix\url{https://link.aps.org/doi/10.1103/PhysRevLett.110.166401}.

\bibitem{Huang}
\bibinfo{author}{Huang, L.}, \bibinfo{author}{Ayral, T.},
  \bibinfo{author}{Biermann, S.} \& \bibinfo{author}{Werner, P.}
\newblock \bibinfo{title}{Extended dynamical mean-field study of the
  $\mathrm{H}$ubbard model with long-range interactions}.
\newblock \emph{\bibinfo{journal}{Phys. Rev. B}} \textbf{\bibinfo{volume}{90}},
  \bibinfo{pages}{195114 -- 195132} (\bibinfo{year}{2014}).
\newblock \urlprefix\url{https://link.aps.org/doi/10.1103/PhysRevB.90.195114}.

\bibitem{Ayral2}
\bibinfo{author}{Ayral, T.}, \bibinfo{author}{Biermann, S.},
  \bibinfo{author}{Werner, P.} \& \bibinfo{author}{Boehnke, L.}
\newblock \bibinfo{title}{Influence of $\mathrm{F}$ock exchange in combined
  many-body perturbation and dynamical mean field theory}.
\newblock \emph{\bibinfo{journal}{Phys. Rev. B}} \textbf{\bibinfo{volume}{95}},
  \bibinfo{pages}{245130 -- 245139} (\bibinfo{year}{2017}).
\newblock \urlprefix\url{https://link.aps.org/doi/10.1103/PhysRevB.95.245130}.

\bibitem{Boehnke}
\bibinfo{author}{Boehnke, L.}, \bibinfo{author}{Nilsson, F.},
  \bibinfo{author}{Aryasetiawan, F.} \& \bibinfo{author}{Werner, P.}
\newblock \bibinfo{title}{When strong correlations become weak: consistent
  merging of $\mathit{GW}$ and $\mathrm{DMFT}$}.
\newblock \emph{\bibinfo{journal}{Phys. Rev. B}} \textbf{\bibinfo{volume}{94}},
  \bibinfo{pages}{201106(R) -- 201110} (\bibinfo{year}{2016}).
\newblock \urlprefix\url{https://link.aps.org/doi/10.1103/PhysRevB.94.201106}.

\bibitem{multitier}
\bibinfo{author}{Nilsson, F.}, \bibinfo{author}{Boehnke, L.},
  \bibinfo{author}{Werner, P.} \& \bibinfo{author}{Aryasetiawan, F.}
\newblock \bibinfo{title}{Multitier self-consistent
  $\mathit{GW}$+$\mathrm{EDMFT}$}.
\newblock \emph{\bibinfo{journal}{Phys. Rev. Materials}}
  \textbf{\bibinfo{volume}{1}}, \bibinfo{pages}{043803 -- 043822}
  (\bibinfo{year}{2017}).
\newblock
  \urlprefix\url{https://link.aps.org/doi/10.1103/PhysRevMaterials.1.043803}.

\bibitem{Golez}
\bibinfo{author}{Gole\ifmmode~\check{z}\else \v{z}\fi{}, D.},
  \bibinfo{author}{Boehnke, L.}, \bibinfo{author}{Strand, H. U.~R.},
  \bibinfo{author}{Eckstein, M.} \& \bibinfo{author}{Werner, P.}
\newblock \bibinfo{title}{Nonequilibrium $\mathit{GW}$+$\mathrm{EDMFT}$:
  antiscreening and inverted populations from nonlocal correlations}.
\newblock \emph{\bibinfo{journal}{Phys. Rev. Lett.}}
  \textbf{\bibinfo{volume}{118}}, \bibinfo{pages}{246402 -- 246407}
  (\bibinfo{year}{2017}).
\newblock
  \urlprefix\url{https://link.aps.org/doi/10.1103/PhysRevLett.118.246402}.

\bibitem{Choi}
\bibinfo{author}{Choi, S.}, \bibinfo{author}{Semon, P.}, \bibinfo{author}{Kang,
  B.}, \bibinfo{author}{Kutepov, A.} \& \bibinfo{author}{Kotliar, G.}
\newblock \bibinfo{title}{$\mathrm{ComDMFT}$: a massively parallel computer
  package for the electronic structure of correlated-electron systems}.
\newblock \emph{\bibinfo{journal}{Computer Physics Communications}}
  \textbf{\bibinfo{volume}{244}}, \bibinfo{pages}{277 -- 294}
  (\bibinfo{year}{2019}).

\bibitem{CTQMC}
\bibinfo{author}{Gull, E.} \emph{et~al.}
\newblock \bibinfo{title}{Continuous-time $\mathrm{Monte}$ $\mathrm{Carlo}$
  methods for quantum impurity models}.
\newblock \emph{\bibinfo{journal}{Rev. Mod. Phys.}}
  \textbf{\bibinfo{volume}{83}}, \bibinfo{pages}{349 -- 404}
  (\bibinfo{year}{2011}).
\newblock \urlprefix\url{https://link.aps.org/doi/10.1103/RevModPhys.83.349}.

\bibitem{CTQMC2}
\bibinfo{author}{Werner, P.} \& \bibinfo{author}{Millis, A.~J.}
\newblock \bibinfo{title}{Dynamical screening in correlated electron
  materials}.
\newblock \emph{\bibinfo{journal}{Phys. Rev. Lett.}}
  \textbf{\bibinfo{volume}{104}}, \bibinfo{pages}{146401 -- 146404}
  (\bibinfo{year}{2010}).
\newblock
  \urlprefix\url{https://link.aps.org/doi/10.1103/PhysRevLett.104.146401}.

\bibitem{space-time}
\bibinfo{author}{Rieger, M.~M.}, \bibinfo{author}{Steinbeck, L.},
  \bibinfo{author}{White, I.}, \bibinfo{author}{Rojas, H.} \&
  \bibinfo{author}{Godby, R.}
\newblock \bibinfo{title}{The $\mathit{GW}$ space-time method for the
  self-energy of large systems}.
\newblock \emph{\bibinfo{journal}{Computer Physics Communications}}
  \textbf{\bibinfo{volume}{117}}, \bibinfo{pages}{211 -- 228}
  (\bibinfo{year}{1999}).
\newblock
  \urlprefix\url{http://www.sciencedirect.com/science/article/pii/S001046559800174X}.

\bibitem{Kutepov}
\bibinfo{author}{Kutepov, A.}, \bibinfo{author}{Haule, K.},
  \bibinfo{author}{Savrasov, S.~Y.} \& \bibinfo{author}{Kotliar, G.}
\newblock \bibinfo{title}{Electronic structure of $\mathrm{Pu}$ and
  $\mathrm{Am}$ metals by self-consistent relativistic $\mathit{GW}$ method}.
\newblock \emph{\bibinfo{journal}{Phys. Rev. B}} \textbf{\bibinfo{volume}{85}},
  \bibinfo{pages}{155129 -- 155150} (\bibinfo{year}{2012}).
\newblock \urlprefix\url{https://link.aps.org/doi/10.1103/PhysRevB.85.155129}.

\end{thebibliography}
\bibliographystyle{naturemag}

\newpage
\begin{figure} [!htbp] 
	\includegraphics[width=0.7\columnwidth, angle=0]{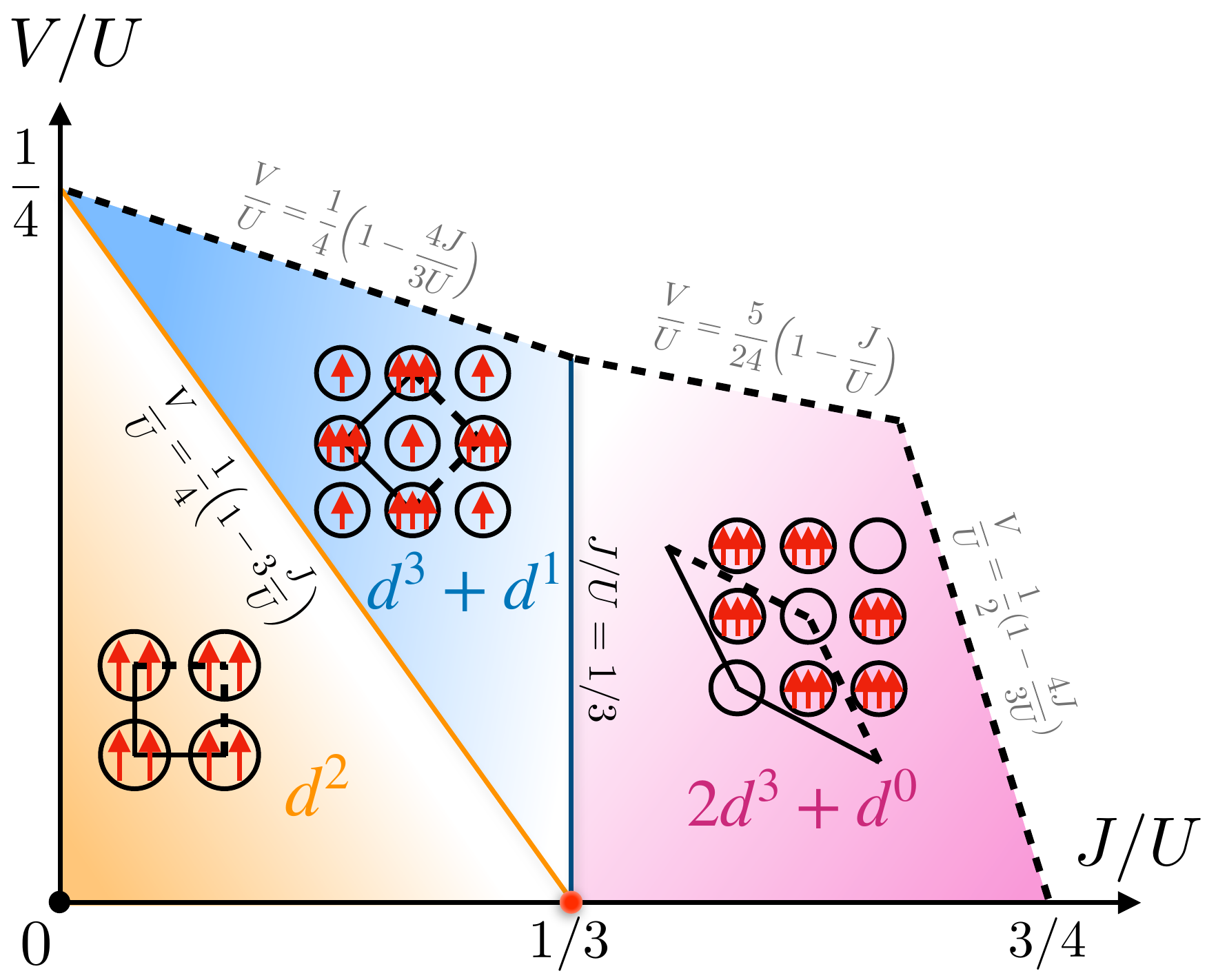}
	\caption{$U$--$V$--$J$ phase diagram obtained at vanishing $t$ and temperature. The lowest energy phases at each $U$, $J$, and $V$ are specified with colored regions; yellow for $d^2$, blue for $d^3+d^1$, and magenta for $2d^3+d^0$. }
	\label{fig1}
\end{figure}

\newpage
\begin{figure} [!htbp] 
	\includegraphics[width=0.7\columnwidth, angle=0]{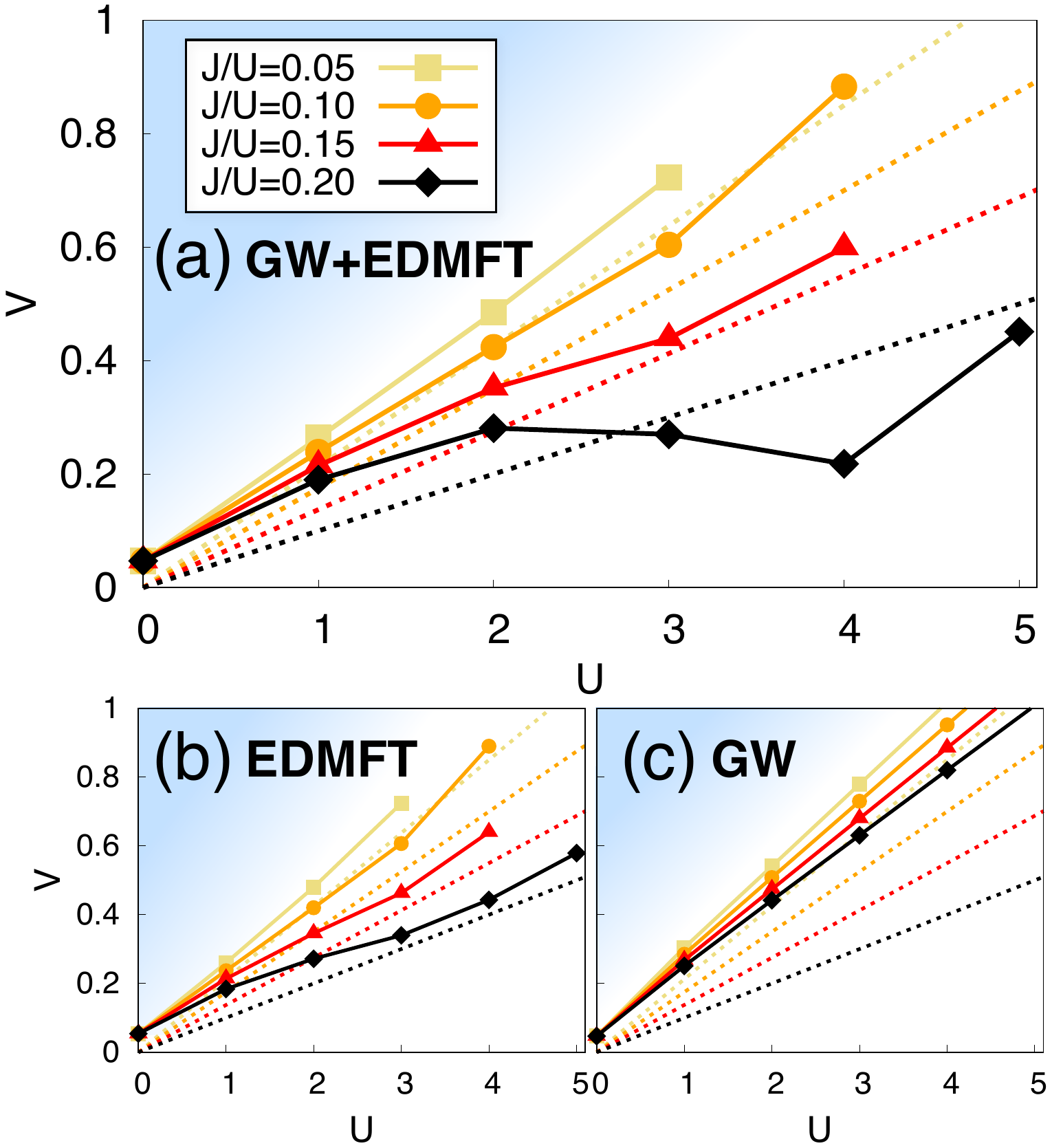}
	\caption{$U$--$V$ phase diagrams with varying $J/U$. The solid lines represents $V_c$ obtained from three different methods: (a) $GW+$EDMFT, (b) EDMFT, and (c) $GW$. The dotted lines represent $V_c$ estimates from the analytical results at atomic limit (see Fig.~\ref{fig1}). The skyblue region highlights the region of $d^3+d^1$ phase. }
	\label{fig2}
\end{figure}

\newpage
\begin{figure} [!htbp] 
	\includegraphics[width=0.7\columnwidth, angle=0]{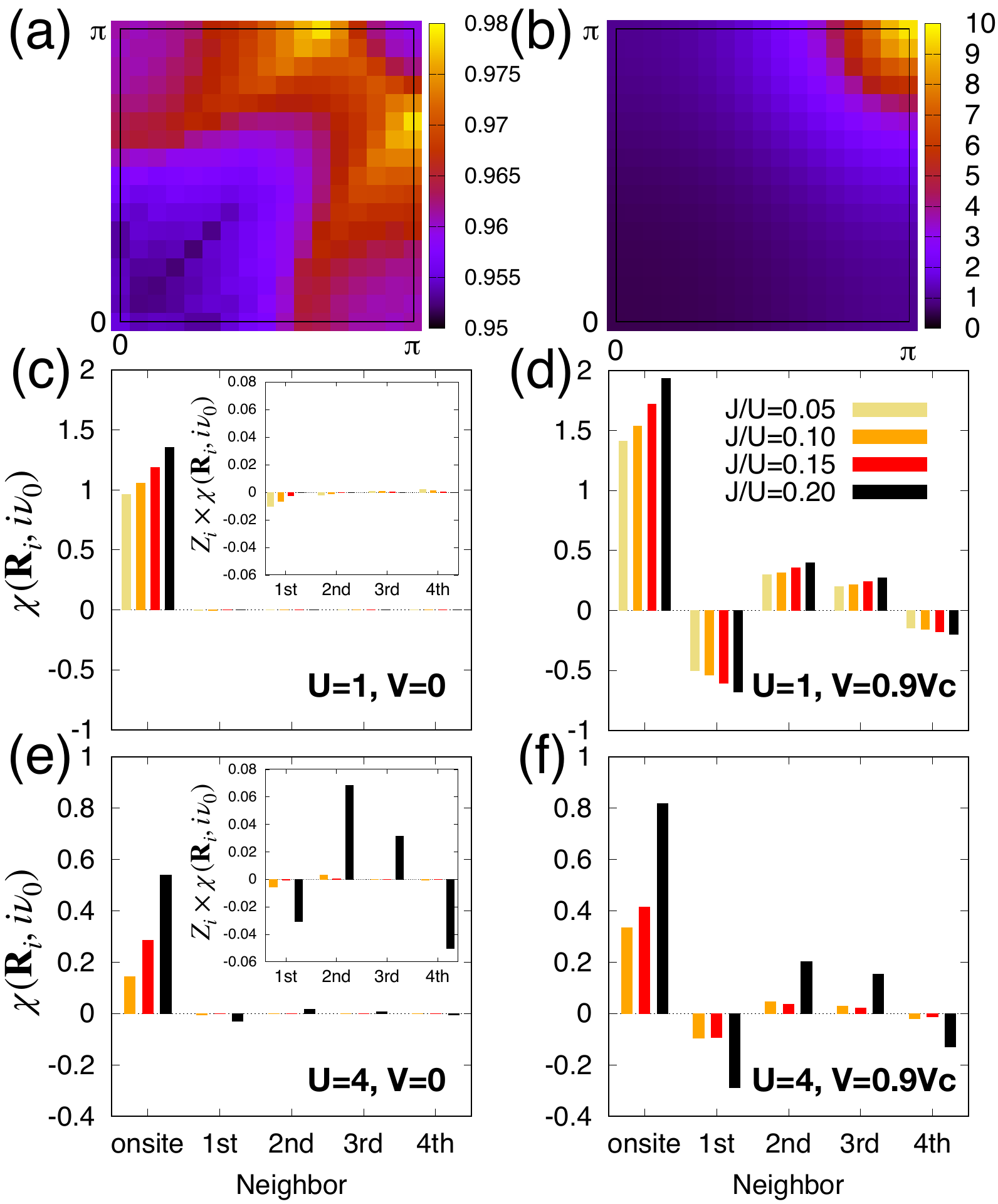}
	\caption{The static charge susceptibility profile in momentum and real spaces. $\chi(\mathbf{k},i\nu_0)$ at (a) $U=1$, $J/U=0.05$, $V=0$ and (b) $U=1$, $J/U=0.05$, $V=0.9V_c$ in the Brillouin zone. (c) -- (f) $\chi(\mathbf{R}_i,i\nu_0)$ as a function of distance. Insets in (c) and (e) highlight the contribution of the coordination number $Z_i$ of the $i$-th NN, and the resulting charge susceptibility.}
	\label{fig3}
\end{figure}

\newpage
\begin{figure} [!htbp] 
	\includegraphics[width=0.7\columnwidth, angle=0]{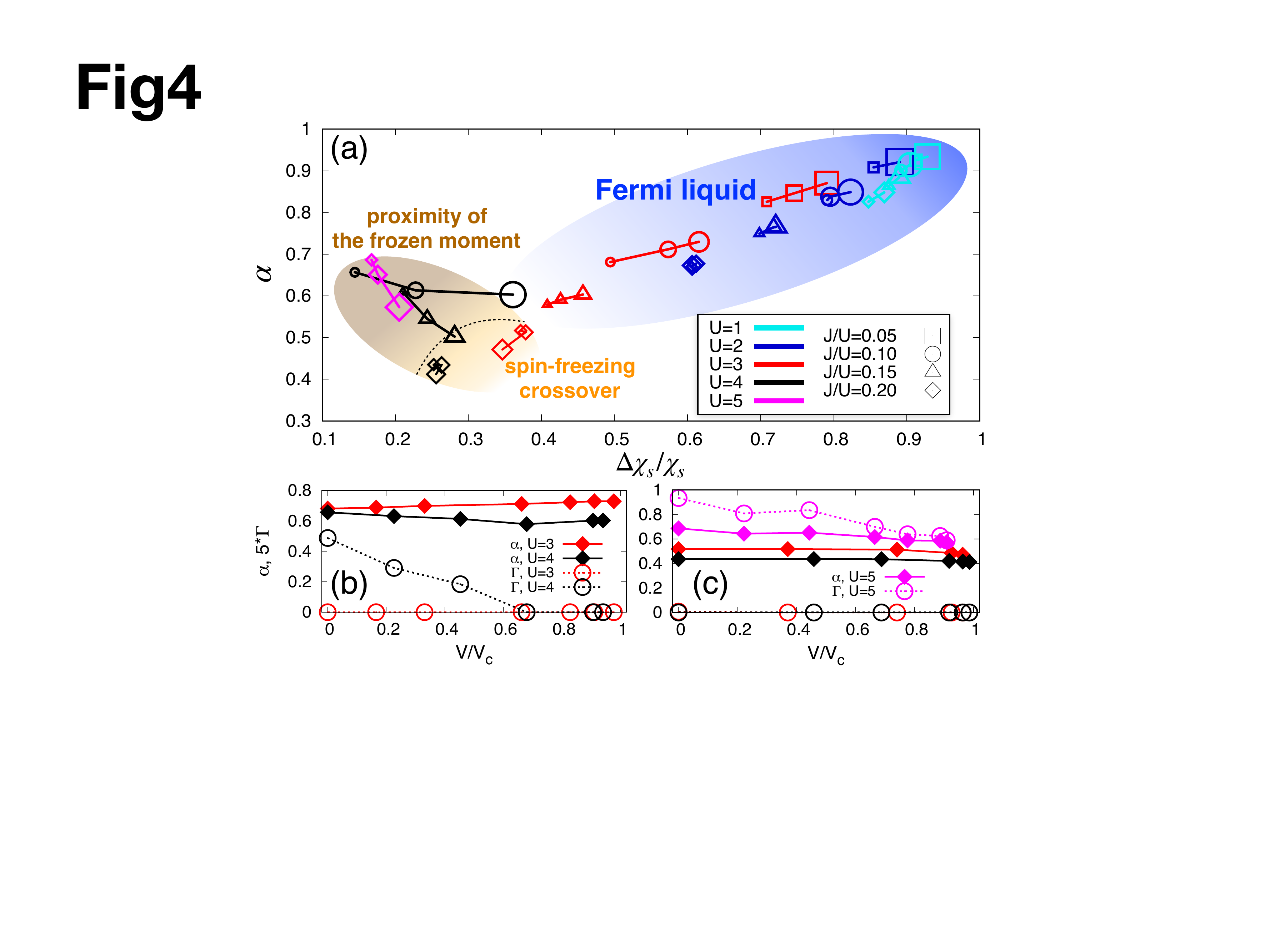}
	\caption{The low frequency behavior of self-energies and their correlation with $\Delta\chi_s/\chi_s$. (a) The correlation between the exponent $\alpha$ and $\Delta\chi_s/\chi_s$. $U$, $J/U$, and $V$ are represented by color, point-shape, and point-size, respectively. Points belonging to the same $U$ and $J/U$ are connected with lines linking from $V=0$ (the smallest point) to the largest $V$ accessible in our numerics (the largest point). The behaviors of $\alpha$ and $\Gamma$ as a function of $V$ are shown for (b) $J/U=0.1$ and (c) $J/U=0.2$. Note that we used three lowest Matsubara frequencies in fitting $\mathrm{Im}\Sigma_\mathrm{loc}(i\omega_n)$ to obtain $\alpha$ and $\Gamma$.}
	\label{fig4}
\end{figure}

\newpage
\begin{figure} [!htbp] 
	\includegraphics[width=0.7\columnwidth, angle=0]{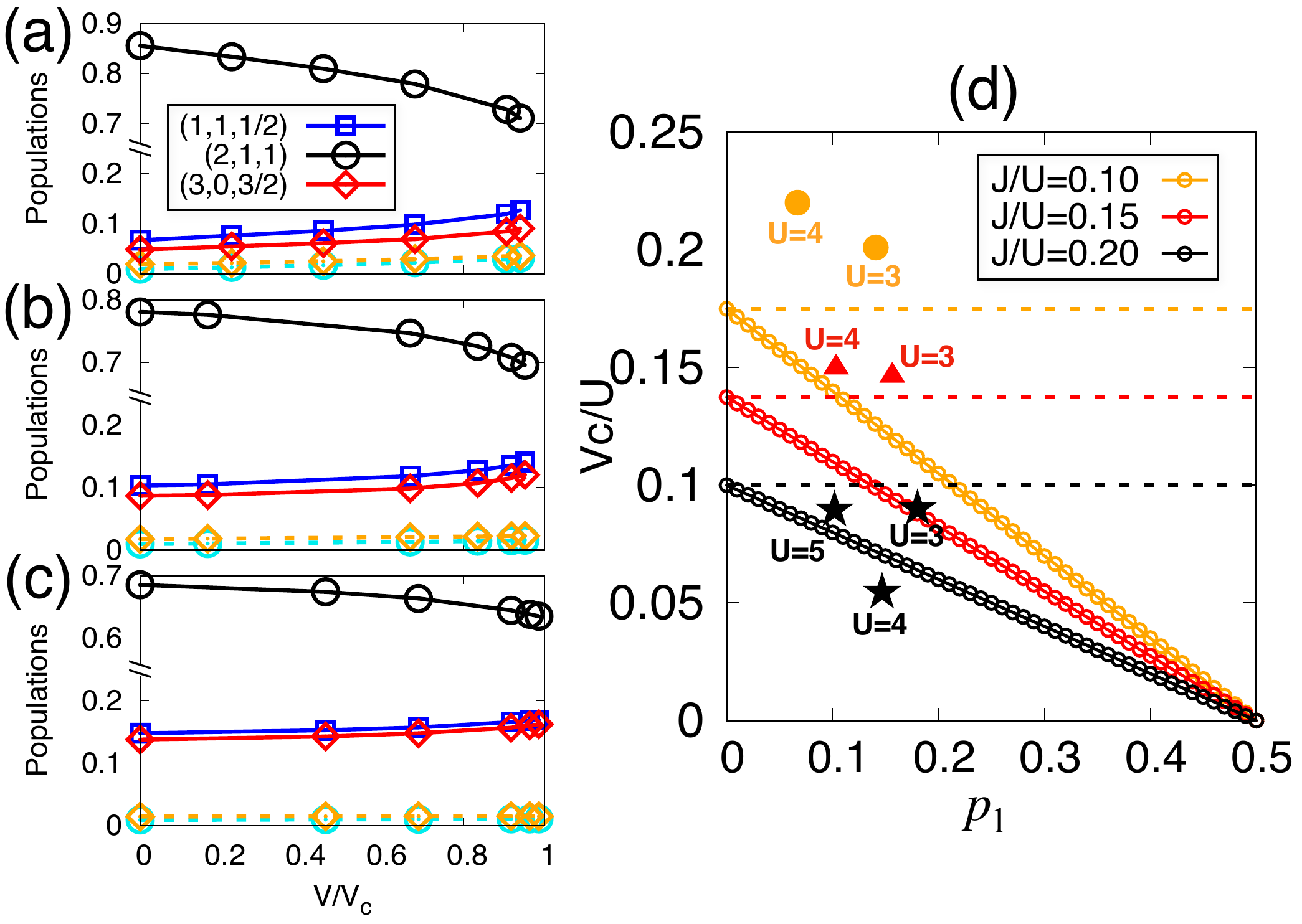}
	\caption{The local multiplet populations and $V_c$ estimates. The local multiplet populations are shown for (a) $U=4$, $J/U=0.1$, (b) $U=4$, $J/U=0.15$, and (c) $U=4$, $J/U=0.2$. Points with solid lines correspond to the maximum $S$ states of each total charge subspace; $(N,L,S) = (1,1,1/2)$, $(2,1,1)$, and $(3,0,3/2)$ indicated by blue (square), black (circle), and red (diamond) lines. Points with dashed lines belong to the remaining smaller $S$ states of $N=2$ (circle) and $N=3$ (diamond) subspaces. (d) $V_c$ estimate as a function of $p_{1}$. The dashed lines correspond to the $V_c$ at $p_1=0$. The actual $V_c$ obtained from $GW+$EDMFT are marked by filled circles ($J/U=0.1$), triangles ($J/U=0.15$), and stars ($J/U=0.2$) (see also Fig.~\ref{fig2}(a)).}
	\label{fig5}
\end{figure}

\clearpage
\onecolumngrid

\setcounter{section}{0}
\setcounter{equation}{0}
\setcounter{figure}{0}

\renewcommand{\thepage}{S\arabic{page}}  
\renewcommand{\thesection}{S\arabic{section}}   

\renewcommand{\theequation}{S\arabic{equation}}
\renewcommand{\figurename}{Supplementary Figure}
\renewcommand{\tablename}{Supplementary Table}

\subsection*{\large
	Supplemental Information for  ``Nonlocal Coulomb interaction and spin-freezing crossover as a route to valence-skipping charge order''
}
\vspace{0.2in}

\subsection*{\large Supplementary Note 1: Atomic limit calculations}
In atomic limit calculations, we restrict ourselves to the integer electron filling on a site with average electron number of $\langle n \rangle=2$ within the unitcell. The Kanamori local interaction Eq.~(\ref{eqS1}) can be rewritten as Eq.~(\ref{eqS2}) \cite{Georges}:	
\begin{align} 
H_\mathrm{loc} &= U\sum_{i,\gamma,\sigma}{n_{i \gamma \uparrow} n_{i \gamma \downarrow}}
+ (U-2J)\sum_{i,\gamma,\gamma'}^{\gamma \neq \gamma'}{ n_{i \gamma \uparrow} n_{i \gamma' \downarrow}} \nonumber 
+ (U-3J)\sum_{i,\gamma,\gamma',\sigma}^{\gamma < \gamma'}{ n_{i \gamma \sigma} n_{i \gamma' \sigma}}  \nonumber \\
&\quad- J\sum_{i,\gamma,\gamma'}^{\gamma \neq \gamma'}{(c^{\dagger}_{i \gamma \uparrow} c_{i \gamma \downarrow} c^{\dagger}_{i \gamma' \downarrow} c_{i \gamma' \uparrow} + c^{\dagger}_{i \gamma \uparrow} c^{\dagger}_{i \gamma \downarrow} c_{i \gamma' \uparrow} c_{i \gamma' \downarrow})}. \label{eqS1} \\
&=(U-3J)\frac{n ( n -1)}{2}+\frac{5}{2}Jn -J(2\vec{S}^2+\frac{1}{2}\vec{L}^2).
\label{eqS2}
\end{align}	
Here $\vec{S}$ and $\vec{L}$ are the spin and orbital angular momentum operators. Thus, the eigenstate of Eq.~(\ref{eqS1}) can be expressed in terms of charge $N$, orbital $L$, and spin $S$. In the case of nonzero and positive Hund's coupling in the atomic limit, the maximum $S$ states are the lowest energy states in each total charge subspace. For corresponding energy levels and related discussions, see Ref.~\cite{Georges}.

Due to the presence nonzero $V$, a specific type of charge ordering pattern is preferred at a given $U$, $J$, and $V$. It thus requires one to search for all ordering patterns to get the lowest energy solution among them. A brute-force numerical approach, however, will easily face the exponentially increasing computational cost. For example, $_{54}\mathrm{C}_{18} \sim \mathcal{O}(10^{13})$ manipulations are required to find out the lowest energy ordering pattern realizable in a periodized $3 \times 3$ cluster. Therefore, for atomic limit calculations presented in the main manuscript, we took nineteen configurations and periodized clusters shown in Supplementary Table~\ref{table_s1}. We compared energy per site of nineteen configurations and all possible ordering patterns thereof.
Although various types of charge orders can emerge depending on the cluster size and configuration, we emphasize that the region of $d^2$ phase and its phase boundary with the valence-skipping $d^3+d^1$ as depicted in Fig.~1 are not affected within our computational setting.

\begin{table} [!htbp] 
	\renewcommand{\arraystretch}{1.3}
	\begin{tabular}{c || c  c  c  c }
		\hline \hline
		cluster size &\ \ $1 \times 1$ &\ \ $\sqrt{2} \times \sqrt{2}$ &\ \ $3 \times 3$ &\ \  $2 \times 2$ and $4 \times 4$ 	\\
		\hline \hline	
		&\ \ $d^2$ &\ \ $\frac{1}{2}d^4+\frac{1}{2}d^0$  &\ \ $\frac{1}{3}d^6+\frac{2}{3}d^0$  &\ \  $\frac{1}{4}d^6+\frac{1}{4}d^2+\frac{1}{2}d^0$ 	 \\
		&\ \ &\ \  $\frac{1}{2}d^3+\frac{1}{2}d^1$   &\ \ $\frac{1}{3}d^5+\frac{1}{3}d^1+\frac{1}{3}d^0$  	&\ \  $\frac{1}{4}d^5+\frac{1}{4}d^3+\frac{1}{2}d^0$  	 \\
		
		&\ \  &\ \   &\ \ $\frac{1}{3}d^4+\frac{1}{3}d^2+\frac{1}{3}d^0$     &\ \  $\frac{1}{2}d^4 + \frac{1}{2}d^0$		 \\
		&\ \ &\ \   &\ \ $\frac{2}{3}d^3+\frac{1}{3}d^0$     &\ \  $\frac{1}{4}d^6+\frac{1}{2}d^1 + \frac{1}{4}d^0$	 	 \\
		
		configuration     &\ \ &\ \   &\ \ $\frac{1}{3}d^4+\frac{2}{3}d^1$      &\ \  $\frac{1}{4}d^5+\frac{1}{4}d^2+\frac{1}{4}d^1+\frac{1}{4}d^0$		 \\
		&\ \ &\ \   &\ \ $\frac{1}{3}d^3+\frac{1}{3}d^2+\frac{1}{3}d^1$     &\ \  $\frac{1}{4}d^4+\frac{1}{2}d^2+\frac{1}{4}d^0$		 \\
		&\ \ &\ \ &\ \ &\ \ $\frac{1}{4}d^5+\frac{3}{4}d^1$ \\
		&\ \ &\ \ &\ \ &\ \ $\frac{1}{4}d^4+\frac{1}{4}d^2+\frac{1}{2}d^1$ \\
		&\ \ &\ \ &\ \ &\ \ $\frac{1}{2}d^3+\frac{1}{2}d^1$ \\
		&\ \ &\ \ &\ \ &\ \ $\frac{1}{4}d^3+\frac{1}{2}d^2+\frac{1}{4}d^1$ \\
		\hline \hline
		
	\end{tabular}
	\caption{Configurations and periodized clusters considered in atomic limit calculations. Here coefficient in front of a valence in a configuration represents the relative proportion of it within the unitcell.}
	\label{table_s1}
\end{table}

\newpage
\subsection*{\large Supplementary Note 2: Computation details for $GW+\mathrm{EDMFT}$ calculations}
The impurity action for $GW+$EDMFT is given by:
\begin{eqnarray}
\begin{split}
S = &-\int_0^\beta d\tau d\tau' \sum_{ab}{c^\dagger_a(\tau) \mathcal{G}_{ab}^{-1}(\tau - \tau') c_b(\tau')}  \\
&+\frac{1}{2}\int_0^\beta  d\tau d\tau' \sum_{abcd}{c^\dagger_a(\tau) c_b(\tau) \mathcal{U}_{abcd}(\tau - \tau') c^\dagger_c(\tau')c_d(\tau')}, 
\end{split}
\end{eqnarray}
where $c^\dagger_a$ ($c_a$) is the electron creation (annihilation) operator for a composite index $a$ labeling both orbital and spin. $\mathcal{G}$ and $\mathcal{U}$ is the fermionic and bosonic Weiss field. The latter is the effective impurity interaction which incorporates the dynamical screening effect. In our case, $\mathcal{U}$ in matsubara frequency space has the form of
\begin{equation}
\mathcal{U}_{abcd}(i\nu_n) = U_{abcd} + D(i\nu_n)\delta_{ab}\delta_{cd},
\end{equation}
which takes into account the dynamical screening $D(i\nu_n)$ in the density-density terms. $U_{abcd}$ is of the Kanamori form.
$\mathcal{G}$ and $\mathcal{U}$ are determined self-consistently in the spirit of EDMFT using impurity self-energy ($\Sigma_\mathrm{imp}$) and polarizability ($P_\mathrm{imp}$):
\begin{eqnarray}
&\mathcal{G}(i\omega_n) = \big(G_\mathrm{loc}^{-1}(i\omega_n)+\Sigma_\mathrm{imp}(i\omega_n)\big)^{-1}, \\
&\mathcal{U}(i\nu_n) = W_\mathrm{loc}(i\nu_n)\big(\mathbf{1}+P_\mathrm{imp}(i\nu_n)W_\mathrm{loc}(i\nu_n)\big)^{-1},
\end{eqnarray}
where $G_\mathrm{loc}$ and $W_\mathrm{loc}$ is the local Green's function and the fully screened Coulomb interaction obtained from Dyson's equations:
\begin{eqnarray}
G_\mathrm{loc}(i\omega_n) = \frac{1}{N_\mathbf{k}}\sum_{\mathbf{k}}{\big(G_0^{-1}(\mathbf{k},i\omega_n)-\Sigma(\mathbf{k},i\omega_n) \big)^{-1}}, \\
W_\mathrm{loc}(i\nu_n) = \frac{1}{N_\mathbf{k}}\sum_{\mathbf{k}}{ \widetilde{V}(\mathbf{k})\big(\mathbf{1}-P(\mathbf{k},i\nu_n)\widetilde{V}(\mathbf{k}) \big)^{-1}}.
\end{eqnarray} 
$G_0$ is the noninteracting Green's function, and $\widetilde{V}(\mathbf{k})$ is the bare Coulomb interaction (including both local and nonlocal terms) in $\mathbf{k}$-space. $N_\mathbf{k}$ is the number of $\mathbf{k}$-points in the first Brillouin zone. In $GW+$EDMFT, $\Sigma$ and $P$ read:
\begin{eqnarray}
\Sigma(\mathbf{k},i\omega_n) = \Sigma_\mathrm{imp}(i\omega_n) + \Sigma^\mathrm{GW}(\mathbf{k},i\omega_n) - \Sigma^\mathrm{GW}_\mathrm{loc}(i\omega_n), \\ 
P(\mathbf{k},i\nu_n) = P_\mathrm{imp}(i\nu_n) + P^\mathrm{GW}(\mathbf{k},i\nu_n) - P^\mathrm{GW}_\mathrm{loc}(i\nu_n).
\label{P}
\end{eqnarray}
In our case, $\chi_\mathrm{imp}$ is of the density-density form and so is the resulting $P_\mathrm{imp}$. The density-density component of the local $GW$ polarizability, $P^\mathrm{GW}_\mathrm{loc}(i\nu_n) = \frac{1}{N_\mathbf{k}}\sum_{\mathbf{k}}P^\mathrm{GW}_{abcd}(\mathbf{k},i\nu_n)\delta_{ab}\delta_{cd}$, is subtracted to avoid the double-counting (Eq.~(\ref{P})).

We performed calculations using $32 \times 32$ $\mathbf{k}$-points in the first Brillouin zone of the square lattice and $\beta =100$. The so-called space-time method \cite{space-time,Kutepov} was applied to calculate the fermionic and bosonic $GW$ quantities, which requires one to evaluate the summation over infinite matsubara frequencies. To this end, we divided the summation interval into two: exact and asymptotic. In the exact interval, we performed the summation exactly. In the asymptotic interval, we replaced $G(\mathbf{k},i\omega_n)$ and $W(\mathbf{k},i\nu_n)$ with their high-frequency analytic forms. 
To ensure a stable convergence, we linearly mixed polarizability between iterations: $P(\mathbf{k},i\nu_n) = (1-R_\mathrm{mix}) P(\mathbf{k},i\nu_n)_\mathrm{old} + R_\mathrm{mix} P(\mathbf{k},i\nu_n)_\mathrm{new}$ with $R_\mathrm{mix} = 0.1$ -- $0.3$ for most cases. We took converged EDMFT solutions as starting points for $GW+$EDMFT calculations.

\subsection*{\large Supplementary Note 3: The charge susceptibility $\chi(\mathbf{k},i\nu_0)$}
\begin{figure*} [!htbp] 
	\includegraphics[width=0.97\columnwidth, angle=0]{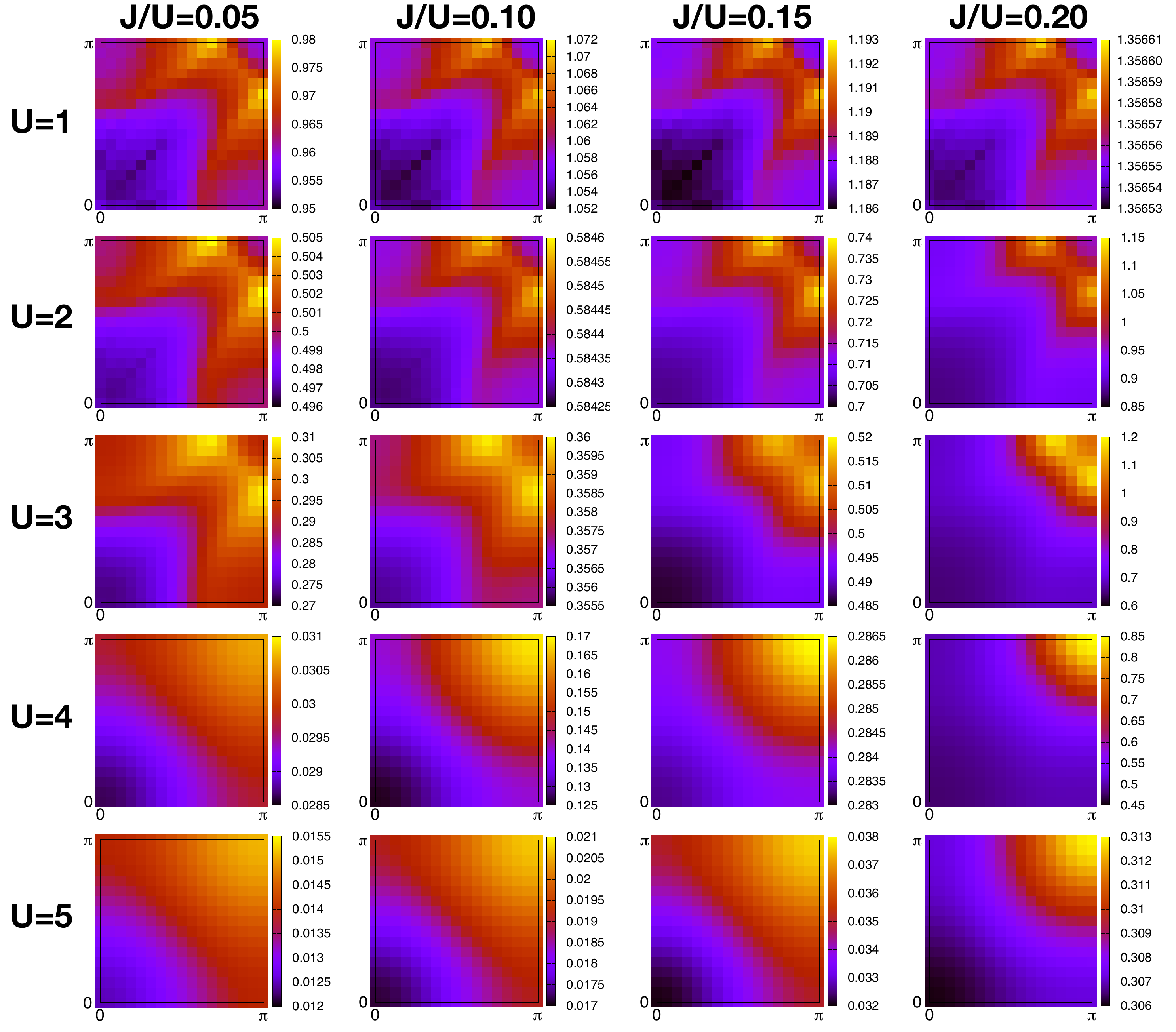}
	\caption{$\chi(\mathbf{k},i\nu_0)$ from $GW+$EDMFT at $V=0$ in the first Brillouin zone.}
	\label{s_fig1}
\end{figure*}

Supplementary Figure~\ref{s_fig1} shows $\chi(\mathbf{k},i\nu_0)$ from $GW+$EDMFT at $V=0$. One can notice that as $U$ increases, $\mathbf{k}$-point where maximum of $\chi(\mathbf{k},i\nu_0)$ appears ($\mathbf{k}_\mathrm{max}$) tends to move from an incommensurate wave-vector to $\mathbf{k}=(\pi,\pi)$. Increasing $J/U$ at fixed $U$ further enhances the magnitude of $\chi(\mathbf{k}_\mathrm{max},i\nu_0)$. Near $V \sim V_c$, $\chi(\mathbf{k},i\nu_0)$ diverges at $\mathbf{k}=(\pi,\pi)$ irrespective of $U$ and $J/U$.

\subsection*{\large Supplementary Note 4: The exponent $\alpha$ and its correlation with $\chi_s^{-1}$ and $\Delta \chi_s$}

In Supplementary Figure~\ref{s_fig2}, we present the correlation of $\alpha$ with $\chi_s^{-1}$ and $\Delta \chi_s$. As discussed in the main manuscript, spin-freezing crossover is in close proximity to the frozen moment regime in which the large $\chi_s$ (or small $\chi_s^{-1}$) emerges. As shown in Supplementary Figure~\ref{s_fig2}(a), spin-freezing behavior is captured within a narrow region of small $\chi_s^{-1}$ ($\chi_s^{-1} \sim 0.05$). 

In Ref.~\cite{Hoshino}, spin-freezing crossover is assigned to a point where the maximum value of $\Delta \chi_s$ is obtained as a function of electron occupation.
The spin-freezing crossover which is identified with $\alpha$ is also found to exhibit large $\Delta \chi_s$ ($\Delta \chi_s \sim 7.5$ -- $8.1$) close enough to the maximum value ($\Delta \chi_s \sim 8.2$) obtained in our parameter range (see Supplementary Figure~\ref{s_fig2}(b)). We note, however, that the spin-freezing is not unambigously identifiable by a clear transition point with $\Delta \chi_s$, but is rather characterized by a broad crossover region around the peak of $\Delta \chi_s$.

\begin{figure*} [!htbp] 
	\includegraphics[width=0.8\columnwidth, angle=0]{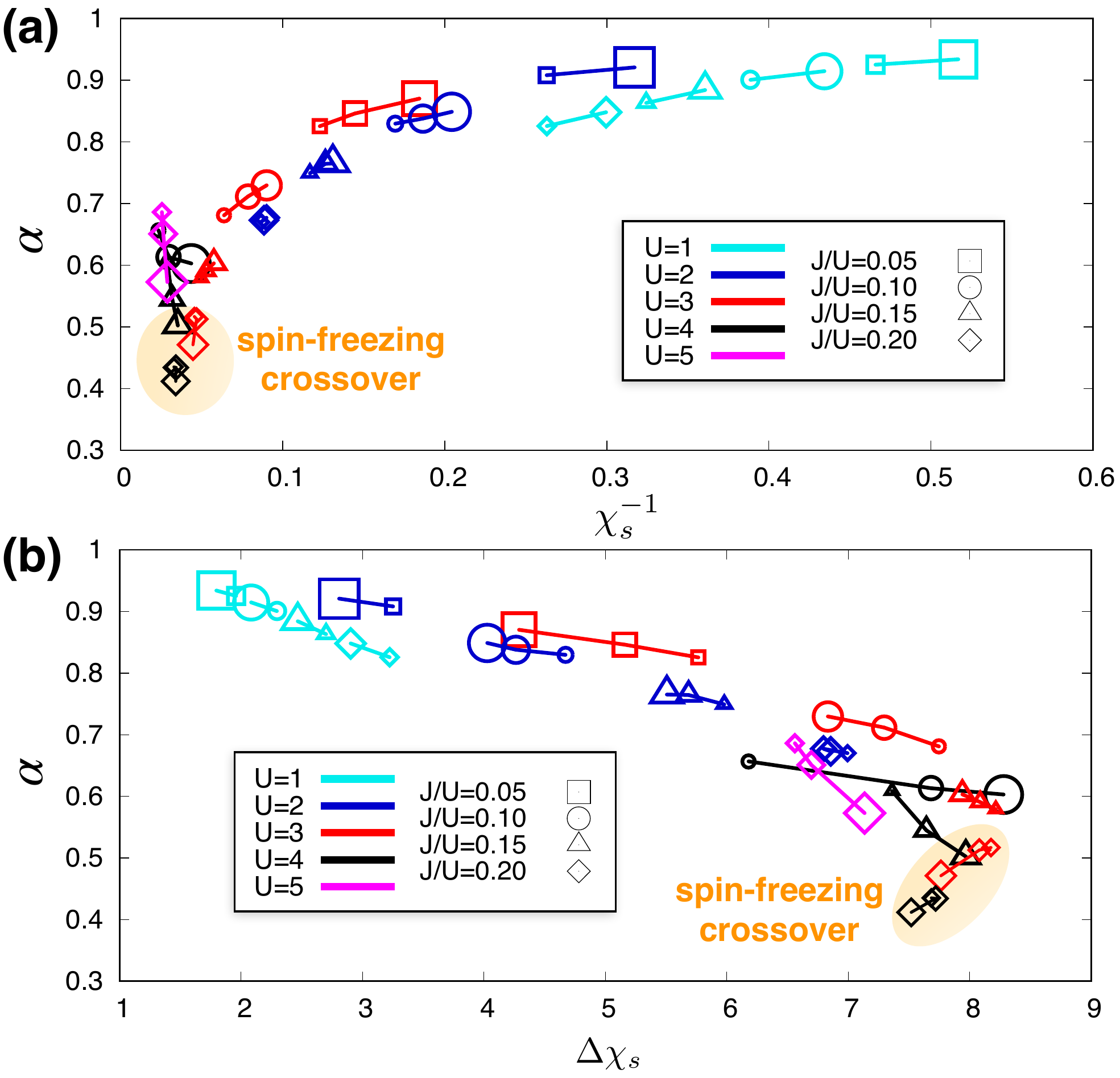}
	\caption{The correlation of the exponent $\alpha$ with (a) $\chi_s^{-1}$ and (b) $\Delta \chi_s$. $U$, $J/U$, and $V$ are represented by color, point-shape, and point-size, respectively. Points belonging to the same $U$ and $J/U$ are connected with lines linking from $V=0$ (the smallest point) to the largest $V$ accessible in our numerics (the largest point). The spin-freezing regime is highlighted by yellow shaded regions.}
	\label{s_fig2}
\end{figure*}

\end{document}